\documentclass{aa}  
\usepackage{graphicx}
\usepackage{txfonts}
\usepackage{booktabs}
\usepackage[breaklinks, colorlinks, citecolor=blue]{hyperref}
\usepackage[normalem]{ulem}
\usepackage{xcolor}

\newcommand{\bperp}{B_\perp}

\defcitealias{Pezzuto2021}{P21}
\defcitealias{Guzman2011}{G11}
\begin{document}

\title{Are Stellar Embryos in Perseus Radio-Synchrotron Emitters?}
\subtitle{Statistical data analysis with {\it Herschel} and LOFAR paving the way for the SKA}

    \authorrunning{Bracco et al.}

   \author{Andrea Bracco\inst{1,2}
          \and
          Marco Padovani\inst{1}
          \and 
          Daniele Galli\inst{1}
          \and 
          Stefania Pezzuto\inst{3}
          \and 
          Alexandre Cipriani\inst{1,2}
          \and
          Alexander Drabent\inst{4}
          }

   \institute{INAF – Osservatorio Astrofisico di Arcetri, Largo E. Fermi 5, 50125 Firenze, Italy\\
   \email{andrea.bracco@inaf.it}
   \and
   Laboratoire de Physique de l'Ecole Normale Sup\'erieure, ENS, Universit\'e PSL, CNRS, Sorbonne Universit\'e, Universit\'e de Paris, F-75005 Paris, France
   \and
   Istituto di Astrofisica e Planetologia Spaziali (IAPS), INAF, Via Fosso del Cavaliere 100, 00133 Roma, Italy 
   \and
   Th{\"u}ringer Landessternwarte, Sternwarte 5, D-07778 Tautenburg, Germany\\
             }

   \date{Received 27 August 2024; accepted 29 November 2024}

 
  \abstract
   {
Cosmic rays (CRs) are crucial to the chemistry and physics of star-forming regions. By controlling the ionization rate of molecular gas, they mediate the interaction between matter flows and interstellar magnetic fields, thereby regulating the entire star-formation process, from the diffuse interstellar medium to the formation of stellar embryos, or cores. The electronic GeV component of CRs is expected to generate non-thermal synchrotron radiation, which should be detectable at radio frequencies across multiple physical scales. However, synchrotron emission from star-forming regions in the Milky Way has been barely observed to date. 

In this work, we present the first attempt to statistically detect synchrotron emission with the LOw Frequency ARray (LOFAR) at 144 MHz from the nearby Perseus molecular cloud (at a distance of $\sim$300 pc). We perform median stacking over 353 prestellar and 132 protostellar cores derived from the {\it Herschel} Gould Belt Survey. Using data from the LOFAR two meter sky survey (LoTSS) with an angular resolution of 20$\arcsec$, we identify 18 potential protostellar candidates and 5 prestellar ones. However, we interpret these as extragalactic contamination in the {\it Herschel} catalog. Our statistical analysis of the remaining cores does not reveal any significant radio counterpart of prestellar and protostellar cores at levels of $5\, \mu$Jy beam$^{-1}$ and $8\, \mu$Jy beam$^{-1}$ in the stacked maps, respectively.

We discuss our non-detections in two ways. For protostellar cores, we believe that strong extinction mechanisms of radio emission, such as free-free absorption and the Razin-Tsytovich effect, may be at play. For prestellar cores, using analytical models of magnetostatic-isothermal cores, the lack of detection with LOFAR helps us constrain the maximum ordered magnetic-field strength statistically attainable by these objects, on the order of 100 $\mu$G.
We predict that the statistical emission of the prestellar-core sample in Perseus as seen by LoTSS should be detectable in only 9 hours and 4 hours with the Square Kilometre Array-Low (SKA-Low) array assemblies AA* and AA4, respectively.
   }


   \maketitle
%
\section{Introduction}\label{sec:intro}

The formation of solar-type stars is determined by highly non-linear physical processes \citep[e.g.,][]{BallesterosParedes2020}. The self-gravity of matter overdensities is contrasted by turbulent and magnetized flows in the interstellar medium (ISM), driving the evolution of diffuse multiphase gas into dense and filamentary molecular clouds where stellar embryos, or cores, originate (see recent reviews by \citealp{Hennebelle2019} and \citealp{Pattle2023}). These cores evolve from gravitationally unstable prestellar cores to main-sequence objects through a series of protostellar phases, during which the core mass progressively accretes from the core envelope onto the central source \citep[e.g.,][]{Andre2000}.

Magnetic fields are believed to regulate this multi-scale flow of matter, from filaments in molecular clouds \citep[e.g.,][]{PlanckXXXII,PlanckXXXV,Soler2019b,Hacar2023} down to the scale of protostellar cores \citep[e.g.,][]{Maury2022}. The magnetic-field strength increases from a few $\mu$G in the diffuse ISM  \citep[for total gas column densities, $N_{\rm H} < 5\times 10^{21}$ cm$^{-2}$,][]{Heiles2003,Thompson2019} to hundreds of mG in denser regions \citep[$N_{\rm H} \sim 10^{24}$ cm$^{-2}$,][]{Crutcher2010}, where most cores appear slightly supercritical with respect to magnetic support, thus inclined toward gravitational collapse \citep[e.g.,][]{Crutcher2012}. 
\begin{figure*}[!ht]
\begin{center}
\resizebox{0.9\hsize}{!}{\includegraphics{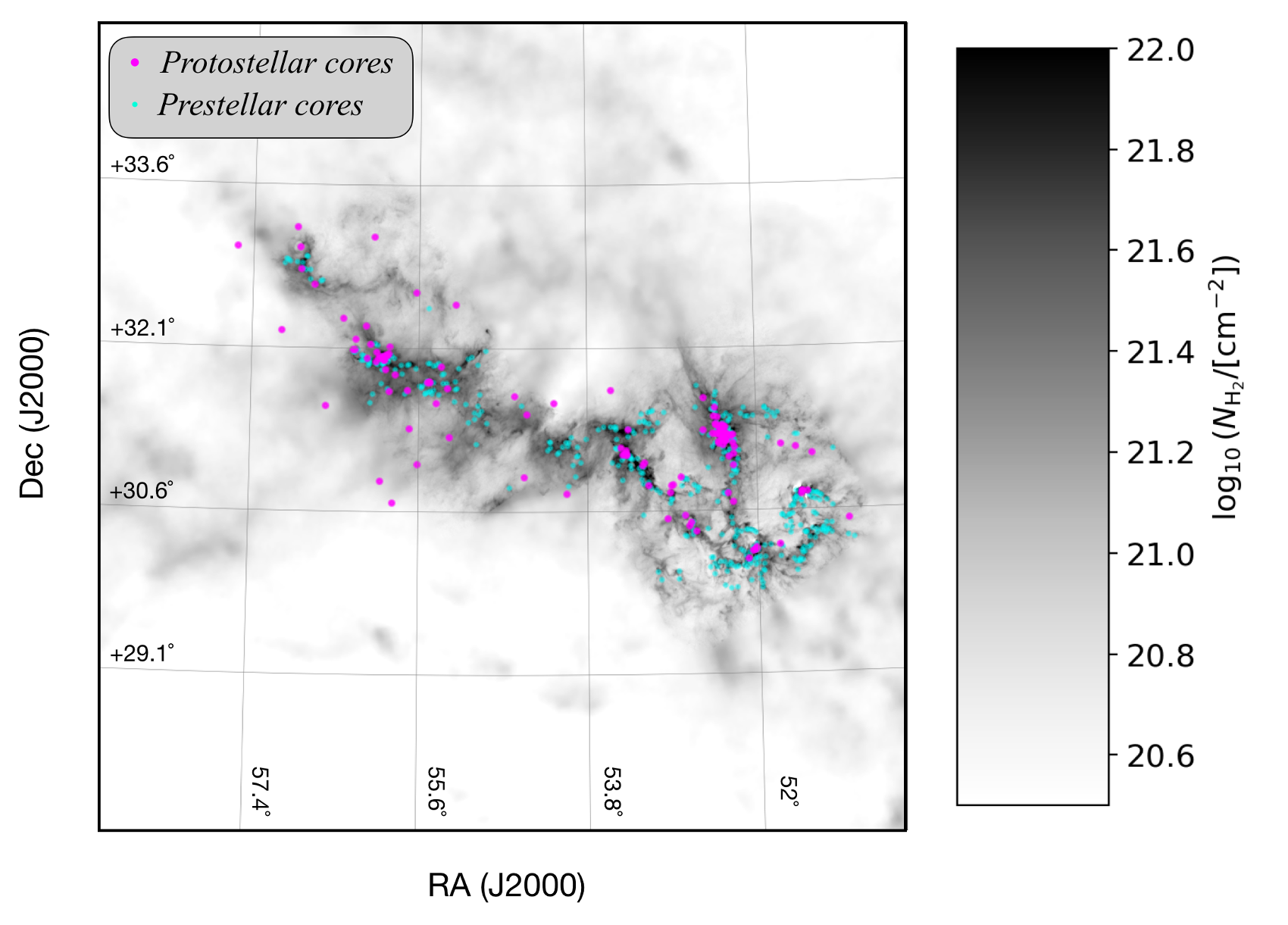}}
\caption{Column density map of molecular hydrogen, $N_{\rm H_2}$, combining {\it Herschel} and {\it Planck} data of the Perseus molecular cloud. Magenta and cyan circles correspond to protostellar and prestellar cores, respectively, as defined in \citetalias{Pezzuto2021}.}
\label{fig:coldens0}
\end{center}
\end{figure*}

The transition to supercritical cores must occur at the expense of some diffusion process that decouples the contrasting effect of magnetic pressure against gravity, such as ambipolar diffusion, which is the relative dynamical drift of ions and neutrals in non-ideal magneto-hydrodynamic (MHD) flows \citep[e.g.,][]{Pinto2008, Momferratos2014}. 
Thus, one key parameter of the star-formation process is the amount of available ionized particles in molecular clouds to link the matter flow with magnetic fields. While ultraviolet (UV) radiation is considered the most powerful ionizing source in the diffuse ISM, cosmic rays (CRs), of either interstellar or local origin, are believed to be the main contributors to the ionization of the dense, UV-shielded portions of molecular clouds \citep[e.g.,][]{Grenier2015, Padovani2020}. CR ionization is primarily determined by the properties of MeV protons \citep[e.g.,][]{Gabici2022}. CR electrons (CRe), while still relevant for the ionization rate of molecular clouds in the MeV range \citep{Ivlev2021,Padovani2022}, are mainly expected to generate non-thermal synchrotron radiation as they gyrate around magnetic-field lines in the GeV range. However, this non-thermal emission from molecular clouds in the Milky Way has been observed in the radio band at low frequencies (below 1 GHz) only in a few instances, such as the diffuse emission in the Orion-Taurus ridge  \citep{Bracco2023} and in the compact regions around the protostellar jet of DG Tau A \citep{FeeneyJohansson2019}.  
The lack of synchrotron emission from molecular clouds represents a long-standing problem that challenges our understanding of CRe and magnetic fields in star-forming regions \citep{Dickinson2015}. 

With the advent of groundbreaking observational facilities, such as the Square Kilometre Array \citep[SKA,][]{Dewdney2009}, we may finally reach the required sensitivity to detect the expected synchrotron emission from individual objects \citep[e.g.,][]{PadovaniGalli2018}. For the first time in this study, we attempt to reveal this emission statistically by performing median stacking over a large sample of prestellar and protostellar cores. In prestellar cores, CRe have an interstellar origin, whereas in protostellar cores, they are expected to be locally accelerated by internal sources, such as jets and shocks \citep{Padovani2015,Padovani2016}. We combine data at 144 MHz from the LOw Frequency ARray \citep[LOFAR,][]{vanHaarlem2013} with the most extensive catalog of cores in the nearby star-forming region of Perseus, as revealed in the infrared by the {\it Herschel} satellite \citep{Pilbratt2010}. 

The paper is organized as follows: in Sect.~\ref{sec:data}, we introduce the data and detail the stacking procedure; in Sect.~\ref{sec:results}, we present our main results on the stacking of LOFAR data over the {\it Herschel} samples of cores; in Sect.~\ref{sec:discussion}, we discuss our results in the context of prestellar and protostellar cores, separately. The paper is summarized in Sect.~\ref{sec:summary} and  contains two appendices (Appendix~\ref{app:mosaic} and Appendix~\ref{app:cutouts}).   
 
\section{Data and Methods}
\label{sec:data}

In this section we describe the {\it Herschel} and LOFAR data used in the analysis and the basic principles of the stacking employed to search for the radio emission of the Perseus cores.   

\subsection{{\it Herschel} data}
\label{ssec:data_description}
\begin{figure*}[!ht]
\begin{center}
\resizebox{1.0\hsize}{!}{\includegraphics{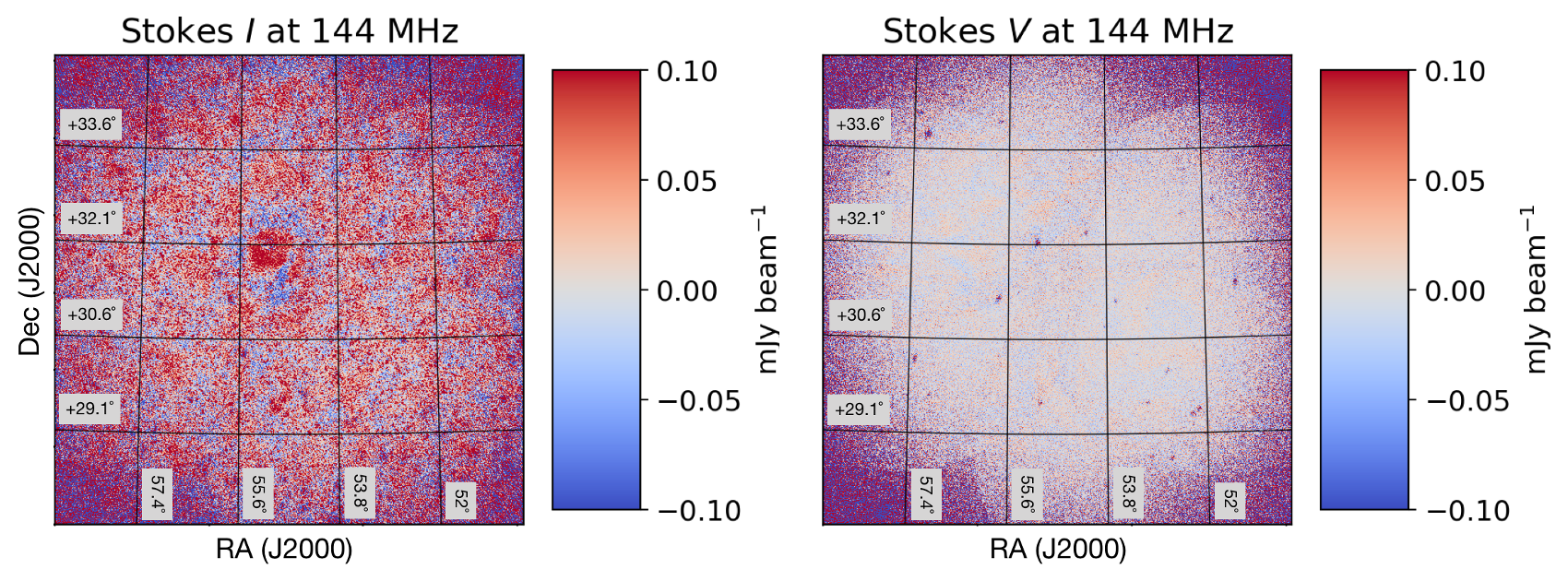}}
\caption{Stokes $I$ and $V$ maps of the Perseus molecular cloud from the LoTSS survey. The maps have the same projection and grid as the $N_{\rm H_2}$ map shown in Fig.~\ref{fig:coldens0}, centered on the celestial coordinate (RA, Dec) (J2000) = (54$^{\circ}$,+31$^{\circ}$).}
\label{fig:lofardata}
\end{center}
\end{figure*}

We considered the catalog of prestellar and protostellar cores in the Perseus molecular cloud released by the {\it Herschel} Gould Belt Survey (HGBS) and detailed in \citet{Pezzuto2021}, hereafter \citetalias{Pezzuto2021}. The cores were identified on the basis of their spectral energy distribution (SED) properties using {\it Herschel} data between 70~$\mu$m and 500~$\mu$m. Two samples of 132 protostellar and 353 prestellar cores 
have been made available via the VizieR data archive\footnote{\url{https://vizier.cds.unistra.fr/viz-bin/VizieR?-source=J/A+A/645/A55}}. We note that, to increase the statistical sample, we included all candidate prestellar cores, even those labeled as tentative detections in the {\it Herschel} catalog. However, we confirmed that our results remain unchanged when these are excluded.  

In Fig.~\ref{fig:coldens0}, we show the location of the full set of cores over the multi-resolution H$_2$ column-density map of the Perseus molecular cloud. This map has a variable angular resolution between 18$\arcsec$ and 5$\arcmin$. It was obtained combining data from the {\it Herschel} and {\it Planck} \citep{PlanckI2016} satellites to recover the diffuse dust emission in the periphery of the HGBS molecular clouds, surveyed at low resolution with {\it Planck}, while preserving the high-resolution dense regions observed by {\it Herschel} \citep[][]{Bracco2020c}. While most of the prestellar cores (cyan) are located within the densest regions of the molecular cloud, protostellar cores (magenta) appear more dispersed, even in the diffuse parts of Perseus. 

\subsection{LOFAR data}\label{ssec:LOFAR}
To look for the meter-wavelength radio emission of the {\it Herschel} cores, we used  20$\arcsec$ resolution Stokes $I$ and $V$ maps at 144 MHz from the latest internal data release of the LOFAR Two-meter Sky Survey \citep[LoTSS,][]{Shimwell2022}. 
LoTSS is a wide area deep radio wavelength sky survey. Most pointings are observed for a total of 8 hrs with 48~MHz (120-168 MHz centered at 144 MHz) of bandwidth which allows for two pointings to be observed simultaneously with current LOFAR capabilities. The data are processed with a direction independent (DI) calibration pipeline that is executed on computer facilities at Forschungszentrum J{\"u}lich and SURF \citep{Drabent2019}.
The DI calibration pipeline used for the data processing is described in \citet{vanWeeren2016} and \citet{Williams2016} and makes use of several software packages including the Default Pre-Processing Pipeline \citep[DP3,][]{vanDiepen2018}, LOFAR SolutionTool \citep[LoSoTo,][]{deGasperin2019} and AOFlagger \citep{Offringa2012}. The pipeline corrects for direction independent errors such as the clock offsets between different stations, ionospheric Faraday rotation, the offset between XX and YY phases and amplitude calibration solutions \citep{deGasperin2019}. The \citet{Scaife2012} flux density scale is used for the amplitude calibration together with the TIFR GMRT Sky Survey alternative data release sky models \citep{Intema2017} of the target fields for an initial phase calibration, although both the amplitude and phase calibration are refined during subsequent processing.

A direction dependent (DD) calibration and imaging
pipeline follows the DI phase. The DD routine makes use of kMS \citep{Tasse2014, Smirnov2015} for direction dependent calibration, and of DDFacet \citep{Tasse2018} to apply the direction dependent
solutions during imaging.  Finally, the restoring beam used in DDFacet for each image product type is kept constant and all image products are made with a uv-minimum baseline of 100m with the uv-maximum baselines varied to provide images at different resolutions -- the highest 6$\arcsec$ resolution images use baselines up to 120km (i.e. all LOFAR stations within the Netherlands). For more details on LoTSS refer to \citet{Shimwell2017}, \citet{Shimwell2019}, and \citet{Shimwell2022}. 

For our purposes, as described in Appendix~\ref{app:mosaic}, we produced a mosaic of five pointings in the surroundings of Perseus reaching a uniform root mean square (RMS) value in Stokes $V$ between 0.1 and 0.2 mJy beam$^{-1}$ in the molecular cloud, where $N_{\rm H_2} > 10^{21}$ cm$^{-2}$ (see Fig.~\ref{fig:rmsV}).
Because of resolved and unresolved (confusion noise) point sources, it is difficult to estimate the noise level in Stokes $I$. On the contrary, Stokes $V$ is considered a good tracer of the thermal noise in radio-interferometric data. In Fig.~\ref{fig:lofardata}, we show the final LOFAR maps used in this work without primary-beam corrections that allow us to access the calibrated fluxes in Stokes $I$. This explains the increase of noise at the edges of the images. The maps are on the same grid as the $N_{\rm H_2}$ map shown in Fig.~\ref{fig:coldens0}. In the Stokes $I$ map there is a noticeable intense circular structure at the center, which is the result of contamination from a bright-source calibration residual.     

\begin{figure*}[!ht]
\begin{center}
\resizebox{.9\hsize}{!}{\includegraphics{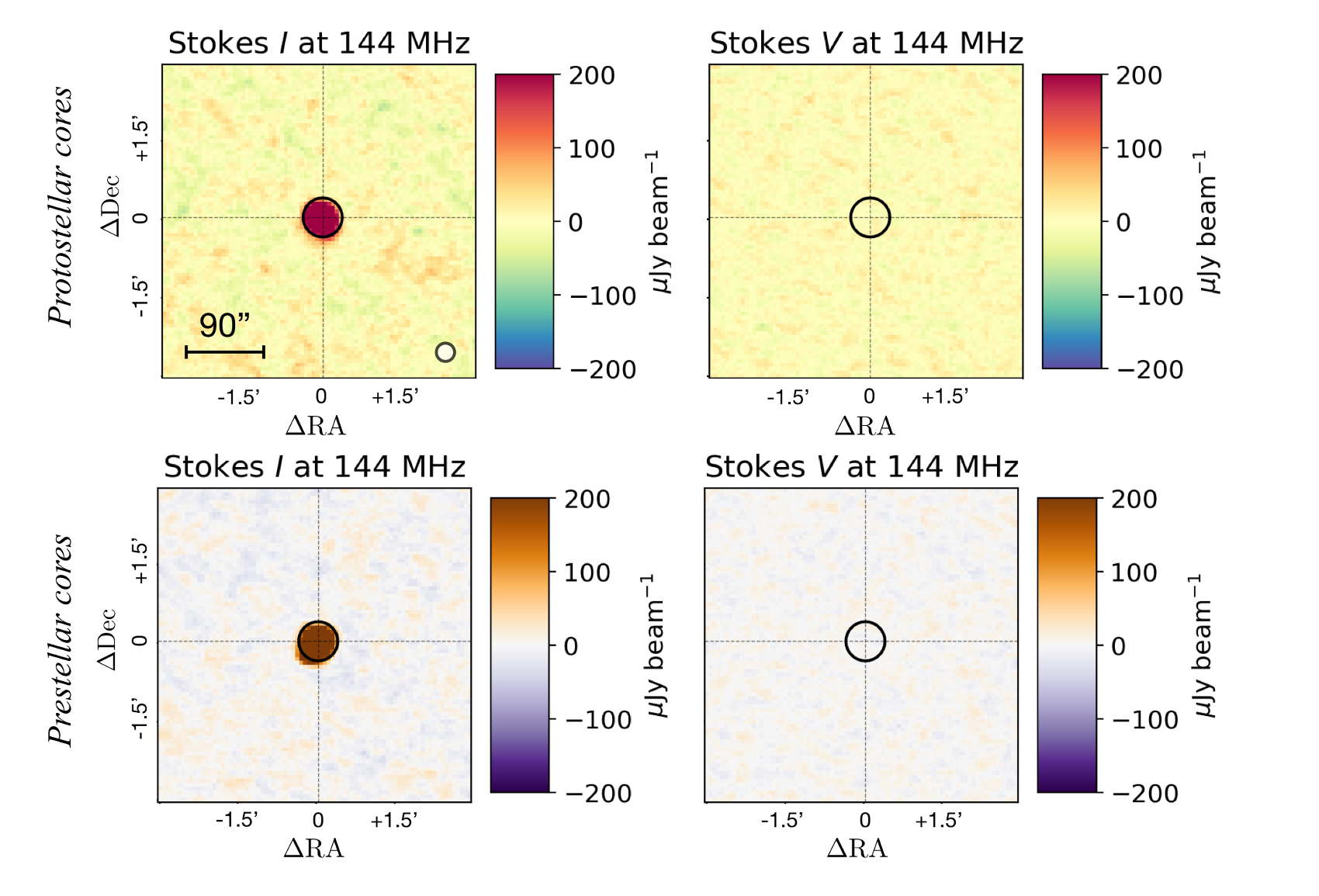}}
\caption{Stacked maps of Stokes $I$ (left) and $V$ (right) from LOFAR at 144 MHz for protostellar (top) and prestellar (bottom) cores in the Perseus molecular cloud. The synthetized beam of 20$\arcsec$-full width at half maximum (FWHM) is shown with a white circle in the top-left panel. The central region where the mean stacking is performed is shown in all panels by the black circle at the center, which is twice the size of the FWHM. At the distance of Perseus ($\sim300$ pc) 90$\arcsec$ correspond to $\sim0.15$ pc.}
\label{fig:stack1}
\end{center}
\end{figure*}

\subsection{Stacking analysis}
\label{ssec:stacking}

We aim to explore whether we can detect LOFAR intensity at the positions of the cores listed in \citetalias{Pezzuto2021}. To this end, we decided to stack Stokes $I$ and $V$ cut-out maps centered on each prestellar and protostellar core separately. While searching for Stokes $I$ counterparts, we used Stokes $V$ as noise reference (see Sect.~\ref{ssec:LOFAR}). The LOFAR maps at hand have a pixel size of 4.5$\arcsec$. Since we do not resolve the cores, we aimed to extract LOFAR intensities within the nominal full width at half maximum (FWHM) of 20$\arcsec$. We produced sufficiently large cut-outs made of 80-pixel-squared maps.

The stacking procedure was twofold. For all cut-out maps outside a circle of 5-pixel radius from the center (roughly twice the size of the FWHM -- see black circle in Fig.~\ref{fig:stack1}), we performed a median stacking, which is less contaminated by strong outlier sources compared to the mean \citep[e.g.,][]{Lee2010}. This choice is justified because outside the center of the stacked maps, we are primarily interested in measuring a meaningful noise level. For the central pixels around the positions of the cores, we instead applied a mean stacking, as we aimed to find correspondence between the {\it Herschel} cores and any bright source detected with LOFAR.

The resulting stacked maps for both prestellar and protostellar cores are shown in Fig.~\ref{fig:stack1}. A black circle at the center separates the central region, where we applied the mean stacking, from the outer region, where we used the median stacking. We note that the median stacking in Stokes $I$ is biased toward positive values compared to Stokes $V$, which fluctuates around zero. This bias, which would be even stronger with mean stacking, is caused by the presence of a few outlier sources at each position in the cut-out maps. The bias is homogeneous across the residual stacked maps, measuring 27 $\mu$Jy beam$^{-1}$ for protostellar cores and 24 $\mu$Jy beam$^{-1}$ for prestellar cores. In the following, we remove the respective biases from the stacked maps.

\section{Results}\label{sec:results}
In this section, we report the main results of our study on the stacked maps, on the source characterization, and on the analysis of the residual stacked maps.

\subsection{Central excess in the stacked Stokes $I$ maps}\label{ssec:stack1}

As shown in Fig.~\ref{fig:stack1}, we detect a clear central excess in the stacked Stokes $I$ maps that is not present in Stokes $V$, both in the case of prestellar and protostellar cores. Since we performed a mean stacking at center, we now explore whether the excess is a statistical property of the {\it Herschel} cores or the contribution of a few bright sources. 
\begin{figure*}[!h]
\begin{center}
\resizebox{1\hsize}{!}{\includegraphics{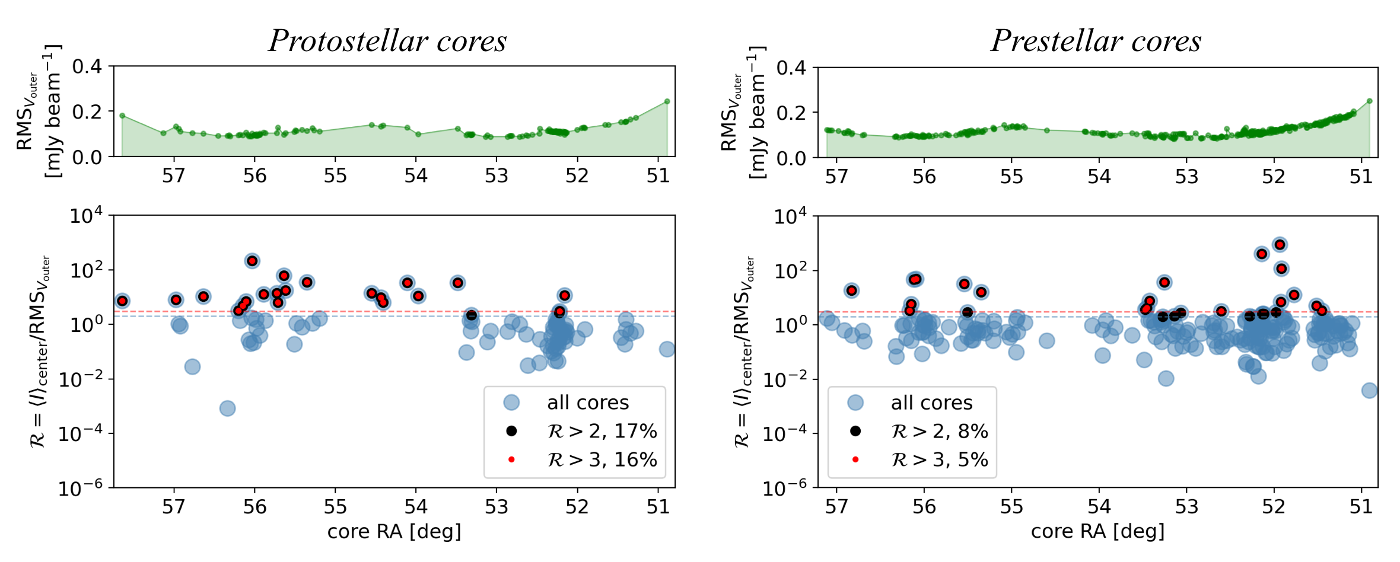}}
\caption{Ratio ($\mathcal{R}$) between the mean of Stokes $I$ within the central region (black circle in Fig.~\ref{fig:stack1}) and the RMS value of Stokes $V$ in the outer region (see top panels), for prestellar (right) and protostellar (left) cores. $\mathcal{R}$ is shown as a function of the RA coordinate of each core. The selected sources correspond to $\mathcal{R}$ values greater than 2 (black dots above the blue dashed-line) and 3 (red dots above the red dashed-line).}
\label{fig:sel}
\end{center}
\end{figure*}

\begin{table*}[ht]
\begin{center}
\caption{Protostellar and prestellar cores with $\mathcal{R} > 3$ and an intensity peak centered on the infrared position (see Figs.~\ref{fig:proto} and \ref{fig:pres}). We include their multi-wavelength counterparts. Using the SIMBAD-CDS astronomical database and the Aladin sky viewer \citep{Boch2014}, optical, near-infrared (NIR), and radio at high frequency (Radio-H) correspond to the observed emission within one LOFAR FWHM (20$\arcsec$) in Pan$-$STARRS1 \citep{Magnier2020}, 2MASS \citep{Skrutskie2006} or Spitzer \citep{Dunham2008,Evans2009}, and VLA \citep{Condon1998, Tobin2016} data, respectively. Detection (non-detection) is referred to with Y (N). The LOFAR column (Radio-L) lists cores with significantly positive values (Y) at the center of each cut-out map. Names and types are from \citetalias{Pezzuto2021}, while indices are used internally in this paper. The label "\ddag" indicates sources present in the NASA/IPAC Extra-galactic Database \citep[NED,][]{Chen2022}. Stars next to types indicate that the sources are absent in SIMBAD-CDS.}
\begin{tabular}{c l c c c c c c c}
\toprule\toprule
Source name & Type & Index & RA (J2000) & Dec (J2000)  & Radio-L & Optical & NIR & Radio-H \\ 
\midrule
HGBS-J032838.0+304008 &  protostellar *, \ddag & 20 & 03:28:38.07 & +30:40:08.1 & Y & Y & Y & Y \\
 HGBS-J033357.1+314329 &  protostellar \ddag & 82 & 03:33:57.15 & +31:43:28.9 & Y & Y & Y & Y \\
HGBS-J033554.3+304500 &  protostellar \ddag & 83 & 03:35:54.35 & +30:45:00.5 & Y & Y & Y & Y \\
HGBS-J033626.3+313639 &  protostellar \ddag & 84 & 03:36:26.34 & +31:36:39.5 & Y & Y & Y & Y \\
HGBS-J033738.3+313009 &  protostellar \ddag & 85 & 03:37:38.37 & +31:30:08.7 & Y& Y & Y & N \\
HGBS-J033745.1+305445 &  protostellar \ddag & 86 & 03:37:45.15 & +30:54:45.3 & Y& Y & Y & N \\
HGBS-J033812.0+314019 &  protostellar \ddag & 87 & 03:38:12.08 & +31:40:18.5 & Y& Y & Y & N \\
HGBS-J034124.9+315656 &  protostellar \ddag & 91 & 03:41:24.98 & +31:56:55.5 & Y& Y & Y & N \\
HGBS-J034227.6+310146 &  protostellar \ddag & 96 & 03:42:27.60 & +31:01:45.7 & Y & Y & Y & Y/N \\
HGBS-J034233.6+323846 &  protostellar *, \ddag & 97 & 03:42:33.63 & +32:38:46.3 & Y& Y & Y & Y \\
HGBS-J034251.1+312216 &  protostellar *, \ddag & 98 & 03:42:51.15 & +31:22:15.8 & Y& Y & Y & N \\
HGBS-J034254.9+314346 &  protostellar \ddag & 99 & 03:42:54.96 & +31:43:45.6 & Y& Y & Y & Y/N \\
HGBS-J034333.3+303958 &  protostellar *, \ddag & 101 & 03:43:33.38 & +30:39:57.9 & Y & N & N & Y \\
HGBS-J034407.6+305227 &  protostellar \ddag & 113 & 03:44:07.64 & +30:52:26.8 & Y & Y & Y & Y \\
HGBS-J034426.0+330950 &  protostellar *, \ddag & 118 & 03:44:26.00 & +33:09:49.6 & Y & Y & Y & N \\
HGBS-J034632.2+313444 &  protostellar \ddag & 126 & 03:46:32.27 & +31:34:43.6& Y & Y & Y & Y/N \\
HGBS-J034754.5+331507 &  protostellar & 130 & 03:47:54.57 & +33:15:06.5 & Y & Y & Y & N \\
HGBS-J035034.9+330400 &  protostellar *, \ddag & 132 & 03:50:34.95 & +33:04:00.1 & Y& Y & Y & N \\
\midrule
HGBS-J032743.8+300716 &  prestellar \ddag & 92 & 03:27:43.89 & +30:07:15.6 & Y & N & N & Y \\
HGBS-J032834.4+305037 &  prestellar *, \ddag & 123 & 03:28:34.42 & +30:50:36.8 & Y & N & N & Y \\
HGBS-J033342.2+311050 &  prestellar *, \ddag & 240 & 03:33:42.22 & +31:10:49.8 & Y & N & N & N \\
HGBS-J034210.8+314153 &  prestellar *, \ddag & 300 & 03:42:10.80 & +31:41:52.5 & Y & N & N & Y \\
HGBS-J034423.1+321001 &  prestellar \ddag & 326 & 03:44:23.10 & +32:10:01.0 & Y & N & N & Y \\
\bottomrule
\label{tab:sources}
\end{tabular}
\end{center}
\end{table*}

For each cut-out map, we computed the ratio  ($\mathcal{R}$) between the mean of Stokes $I$ within the central region and the corresponding RMS value of Stokes $V$ in the outer region. The resulting $\mathcal{R}$ values are shown in Fig.~\ref{fig:sel} for both prestellar and protostellar cores as a function of the RA coordinate of each core in degrees. In the top panels of the figure, we also show the corresponding RMS values of $V$, which are consistent with the RMS values of the mosaic as described in Sect.~\ref{ssec:LOFAR} and shown in Fig.~\ref{fig:rmsV}. 

For both types of cores we define as outliers those with $\mathcal{R} > 3$ (red-dashed line), or the 16\% and the 5\% of the protostellar and prestellar cores, respectively. To verify the robustness of this threshold, in the following we will also consider the case with $\mathcal{R} > 2$ (blue-dashed line), or the 18\% and the 7\% of the protostellar and prestellar core samples. All outlier-maps with $\mathcal{R} > 3$ can be seen in Figs.~\ref{fig:proto} and \ref{fig:pres}. By selecting the outliers with intensity peaks at the center of each cut-out, we detected 18 protostellar cores and 5 prestellar cores that are listed in Table~\ref{tab:sources}, or 14\% and 1\% of the corresponding samples of cores. 

\subsection{Characterization of the selected bright cores}\label{ssec:sources}

In order to assess the robustness of the correlation between these 23 {\it Herschel} cores with LOFAR against the contamination of extragalactic sources, we searched the literature for multi-wavelength catalogs. Using the SIMBAD-CDS astronomical database and the Aladin sky viewer \citep{Boch2014}, we looked for emission in the optical with Pan-STARRS1 \citep{Magnier2020}, in the near-infrared (NIR) with 2MASS \citep{Skrutskie2006} and Spitzer \citep{Dunham2008,Evans2009}, and in the cm-radio range with the VLA \citep{Condon1998, Tobin2016}. We also inquired the NASA/IPAC Extra-galactic Database \citep[NED,][]{Chen2022}.

In Table~\ref{tab:sources}, we list the selected protostellar and prestellar cores as reported in \citetalias{Pezzuto2021}, along with their names and locations. We also specify their indices (internal to this work) and their multi-wavelength counterparts. If a core is detected within the LOFAR FWHM in the other datasets, it is marked with a "Y"; if not detected, it is marked with an "N". Uncertain cases are indicated with "Y/N". For each core, the star symbol "*" indicates that the core is absent as a known source in SIMBAD-CDS, and the label "\ddag" indicates that it is present in NED. 

All protostellar cores with detected LOFAR emission at the center are listed in NED, with the closest counterpart being within only 5$\arcsec$. These cores also emit in the optical, NIR, and cm-radio ranges. However, aside from a few cases with clear extended emission (e.g., indices 97 and 113), which are likely external galaxies, it is challenging to determine whether the origin of the remaining cores is Galactic or extragalactic. The cores without the "*"-symbol are cataloged as young stellar objects, although with low confidence level and high likelihood to be contaminating galaxies \citep[section 5.2 in][]{Dunham2008}. These evidences lead us to surmise that the LOFAR excess corresponding to the protostellar cores detected by \citetalias{Pezzuto2021} is generally the result of background emission from external galaxies.

\begin{figure*}[!h]
\begin{center}
\resizebox{0.9\hsize}{!}{\includegraphics{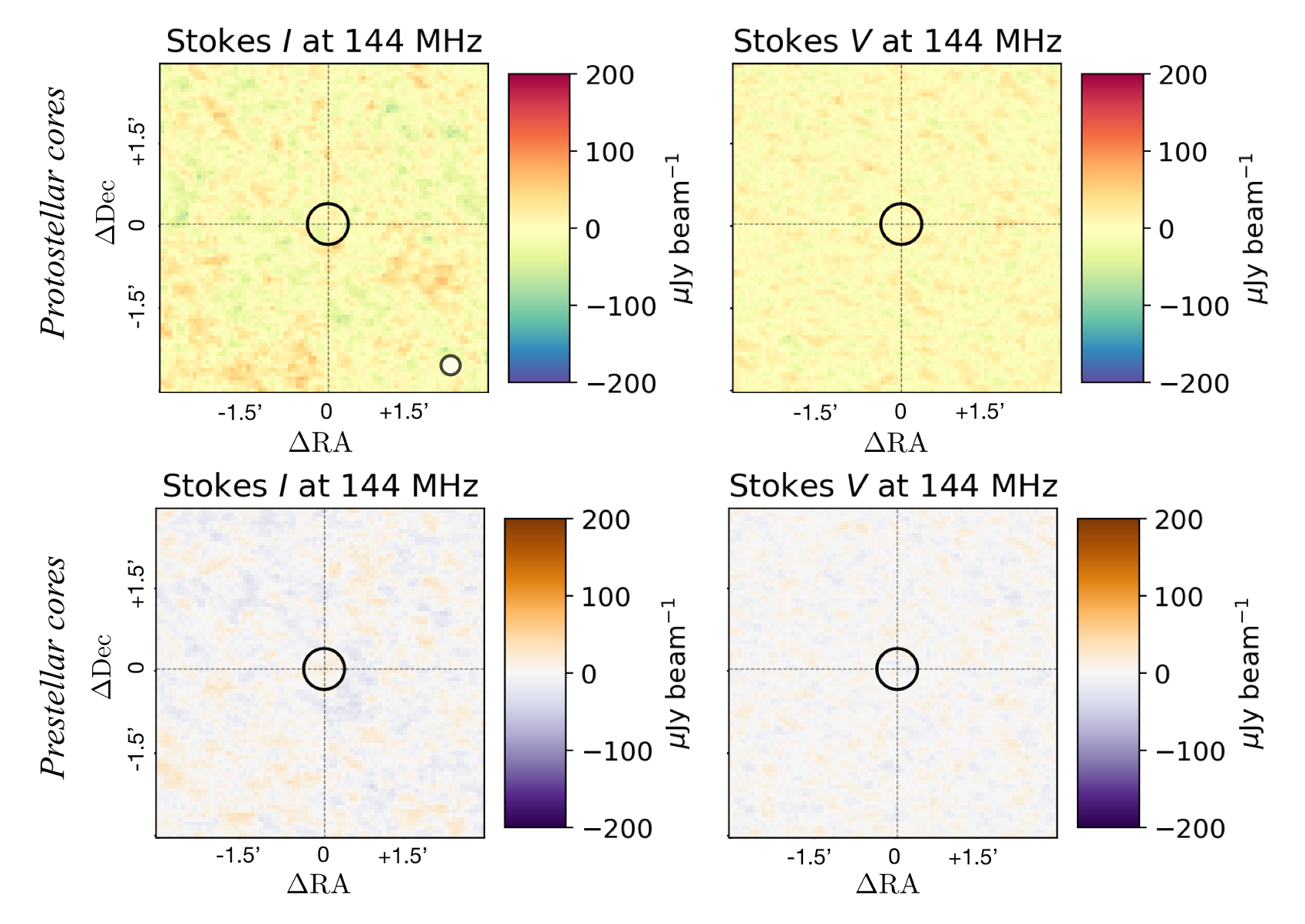}}
\caption{Residual stacked maps without the selected sources shown in Fig.~\ref{fig:sel}, \ref{fig:proto}, and \ref{fig:pres}.}
\label{fig:stack2}
\end{center}
\end{figure*}

\begin{figure*}[!h]
\begin{center}
\resizebox{1\hsize}{!}{\includegraphics{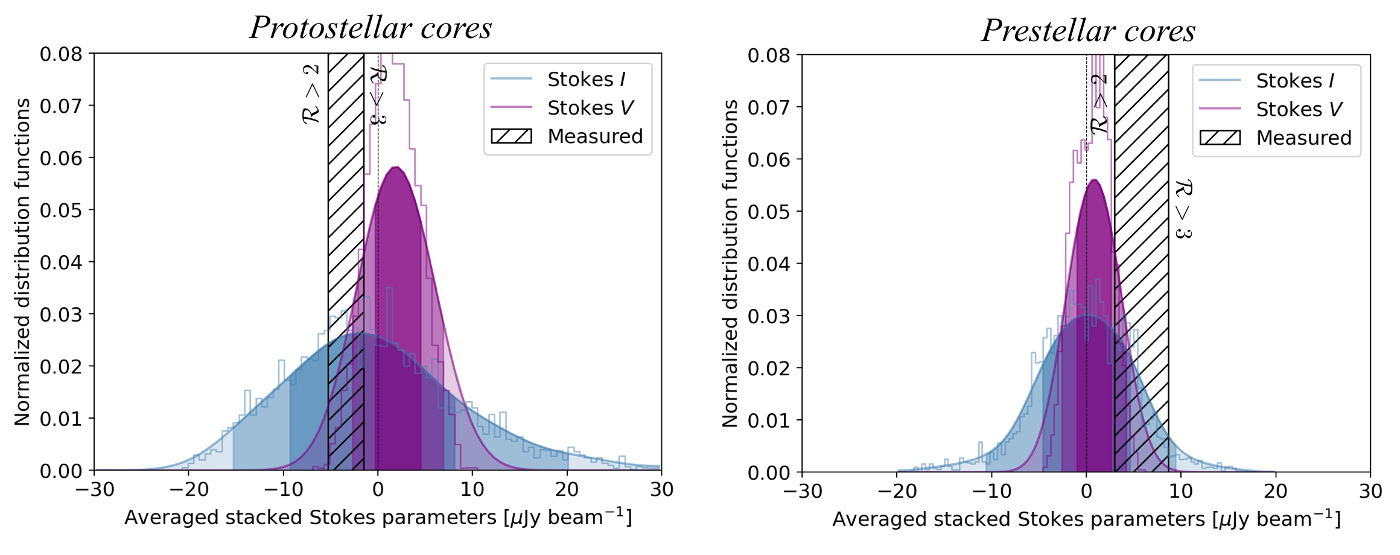}}
\caption{Normalized distribution functions (NDF) for protostellar (left) and prestellar (right) cores of the residual stacked Stokes-$I$ (blue) and -$V$ (purple) maps averaged within 5000 circles of equal area as the central black circle in Figs.~\ref{fig:stack1} and \ref{fig:stack2}. The measured values at the center are shown with black-diagonal hatches depending on the choice of  $\mathcal{R}$ for defining the outlier cores. From dark to light colours, shadows correspond to 1-, 2-, and 3-$\sigma$ levels. The 84-th percentiles are 8 $\mu$Jy beam$^{-1}$ and 5 $\mu$Jy beam$^{-1}$ for protostellar and prestellar cores, respectively. The NDF are shown with both histograms and kernel density estimates.}
\label{fig:distributions}
\end{center}
\end{figure*}

In the case of prestellar cores, the correlation between the {\it Herschel} and LOFAR data is even poorer than in the protostellar case. Some of them, as shown in Fig.~\ref{fig:pres}, are likely signatures of jets from radio galaxies given their extended bipolar morphology. Additionally, despite not emitting either NIR or optical radiations, as expected in the case of prestellar cores, the sample of 5 candidate cores is present in NED. 
Finally, as will be discussed in Sect.~\ref{ssec:prestellar}, the current theoretical understanding of non-thermal synchrotron radiation in prestellar cores struggles to reproduce the mJy beam$^{-1}$ levels observed in LOFAR. For all these reasons, we conclude that the central excess found in prestellar cores is due to contamination from external galaxies. Finally, We verified that all sources listed as galaxies in \citetalias{Pezzuto2021} (see their Table E.1) exhibit a central excess in LOFAR Stokes $I$ at a level of mJy beam$^{-1}$. 

\subsection{Residual stacked maps}
\label{ssec:residu}

We repeated the stacking analysis after excluding the contaminating extragalactic sources from the core catalog. In Fig.~\ref{fig:stack2}, we show the residual stacked maps for both prestellar and protostellar cores, where the central excess is now removed for both types of cores. We obtained  residual stacked maps of Stokes $I$ that closely resemble noise if compared to Stokes $V$.   
This is supported by comparing the normalized distribution functions (NDF) of the residual stacked maps. In Fig.~\ref{fig:distributions}, we show the NDF -- both as histograms and as kernel density estimates (KDE) -- of the averaged Stokes $I$ (in blue) and Stokes $V$ (in purple) within 5000 circles of equal area to the central region delimited by the black circle in, for instance, Figs.~\ref{fig:stack1} and \ref{fig:stack2}. These circles are uniform-randomly distributed over the maps. As expected in case of noise, the NDFs are consistent with zero within 1-$\sigma$ for both prestellar and protostellar cores, and for both Stokes $I$ and Stokes $V$ residual stacked maps. Since the sample of prestellar cores is much larger than that of protostellar cores (348 versus 114), the corresponding NDFs are generally narrower. The 84-th percentile values of the Stokes $I$ NDFs are 5 $\mu$Jy beam$^{-1}$ for prestellar cores and 8 $\mu$Jy beam$^{-1}$ for protostellar cores.
We also note that the NDF of Stokes $I$ is broader than that of Stokes $V$, as Stokes $I$ is more sensitive to calibration uncertainties -- because of the large amount of point sources -- than Stokes $V$, which should is dominated by thermal noise \citep{Shimwell2022}.

Finally, in Fig.~\ref{fig:distributions}, we show with black diagonal hatches the range of average values of the residual stacked maps within the central region both considering $\mathcal{R} >2$ and > 3. While for the protostellar case this average is well within 1-$\sigma$ of the corresponding NDF, for prestellar cores, the case with $\mathcal{R} > 3$ slightly falls above the 84-th percentile. However, the significance of this result is marginal, given the stronger influence of outlier sources in the mean stacking applied to the central region compared to the median stacking used in the outer region, as shown by the case with $\mathcal{R} > 3$. 
Overall, our statistical analysis fails to detect any significant emissions from prestellar and protostellar cores at meter wavelengths.

\section{Discussion}
\label{sec:discussion}

In this section, we investigate several scenarios that might explain why no statistical signature of stellar embryos in Perseus was found using LOFAR. We divide the discussion into two parts: protostellar cores (Sect.~\ref{ssec:ext}) and prestellar cores (Sect.~\ref{ssec:prestellar}). 

\subsection{Extinction mechanisms of protostellar synchrotron emission}
\label{ssec:ext}

Non-thermal synchrotron emission from protostellar cores is expected via two main mechanisms: shocks at the surface of protostars and shocks 
associated to protostellar jets \citep[e.g.,][]{Padovani2015, Padovani2016, Gaches2018, Padovani2021b}. Observational evidence of these processes has been primarily shown in the radio at cm-wavelengths \citep[e.g.,][]{Anglada1995, Rodriguez2016, Rodriguez2017, Purser2018, Sanna2019}, where, nonetheless, the non-thermal radiation is intertwined with thermal components primarily caused by ionizing jets \citep[][]{Tobin2016, Purser2018, Tychoniec2018}.

One case of detected radio emission was reported with LOFAR toward the T Tau young stellar object in Taurus, which was attributed, however, to free-free thermal emission  \citep{Coughlan2017}.
Examples of non-thermal emission from protostellar cores at meter wavelengths are rare. The only case known to us is the LOFAR detection of two knots of synchrotron emission along the jet of DG Tau A on a 1000 au scale in the Taurus molecular cloud \citep[$\sim140$ pc,][]{FeeneyJohansson2019}. Given the LOFAR FWHM, such a signal would likely be averaged within one resolution element at the distance of Perseus \citep[$\sim300$ pc,][]{Zucker2019,Doi2021}. The lack of any statistical signal of radio emission from our sample of protostellar cores with LOFAR at 144 MHz in the Perseus cloud suggests that potential extinction mechanisms may be at play. This interpretation is even more significant, as no synchrotron radiation has been detected from well-known protostellar core systems in Perseus, which have estimated magnetic field strengths of hundreds of $\mu$G, such as NGC 1333 IRAS4 \citep{Girart2006} or Per-B1 \citep{Coude2019}. 

\begin{figure}[!t]
\begin{center}
\resizebox{1.0\hsize}{!}{\includegraphics{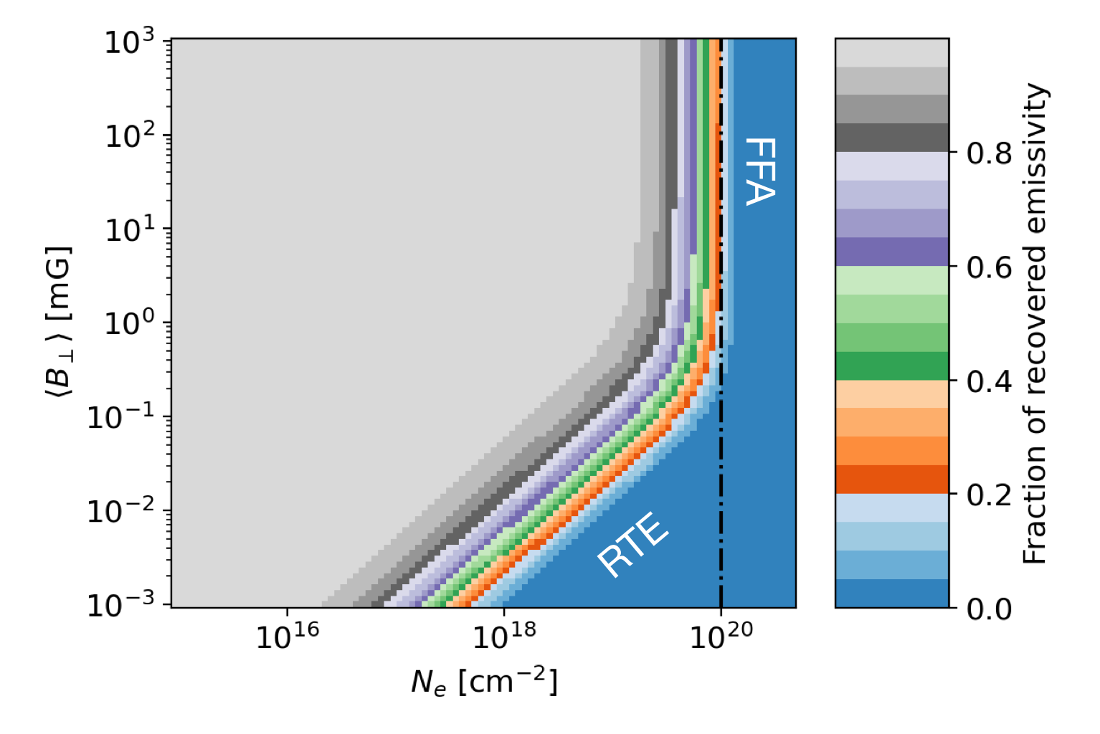}}
\caption{Fraction of modeled synchrotron emissivity of protostars recovered after accounting for synchrotron self absorption, free-free absorption (FFA) and Razin-Tsytovich effect (RTE). The fractional emissivity at 144 MHz is shown as a function of $\langle B_{\perp}\rangle$ and $N_e$. The temperature of the ionized gas causing FFA is considered $10^4$ K. The extinction processes are averaged within 0.03 pc from the modeled protostar, or the LOFAR FWHM at the distance of Perseus. A vertical black dashed-dotted line marks the estimated $N_e$ value for our protostellar-core sample.}
\label{fig:ext}
\end{center}
\end{figure}

Non-thermal radiation in the radio at long wavelengths can be affected by three main processes \citep{Ginzburg1965}: synchrotron self-absorption (SSA), free-free absorption (FFA), and the Razin-Tsytovich effect (RTE). While SSA becomes significant only at frequencies below 10 MHz with standard magnetic field strengths of protostellar cores \citep[from hundreds of $\mu$G to mG; see also][]{FeeneyJohansson2019}, FFA and RTE can have a dramatic effect at LOFAR frequencies.

We modeled the synchrotron emissivity that would be produced in the protostellar case within the LOFAR FWHM at the distance of Perseus, including both FFA and RTE on the same scales. For the synchrotron emissivity, $\varepsilon_{\nu}$, we used Equations (1)--(5) in \citet{Padovani2021}, assuming the CRe energy spectrum $j_e(E)$ from \citet{Bracco2024}, which reproduces the observed radio spectral indices in the literature below 408 MHz. For modeling FFA and RTE, we followed \citet{Stanislavsky2023}, \citet{Bracco2023}, and \citet{Ginzburg1965}.
FFA is an absorption effect of radio photons due to thermal charged particles along the sight line. We computed the corresponding optical depth $\tau_{\nu,\mathrm{FFA}}$ responsible for the absorption $e^{-\tau_{\nu,\mathrm{FFA}}}$ as, 
\begin{equation}\label{eq:tauff}
    \tau_{\nu,\mathrm{FFA}} = 3.014\times10^4\, Z \left( \frac{T_e}{{\rm K}} \right) ^{-1.5}\left( \frac{\nu}{{\rm MHz}} \right)^{-2} \left(\frac{EM}{{\rm cm^{-6}\, pc}}\right) g_{\rm ff},
\end{equation}
where $Z$ is the average number of ion charges assumed 2.5 \citep[including a small contribution from heavier elements than hydrogen,][]{Stanislavsky2023}, $T_e$ is the electron temperature, and $EM$ is the emission measure, that is, the integral along the line of sight (LoS) of the electron density, $n_e$, squared (i.e., $EM = \int_{\rm LoS} n^2_{e}{\rm d}l$). The $g_{\rm ff}$ value is the Gaunt factor equal to $\ln{(49.55/Z/\nu)} + 1.5\ln{T_e}$, where $\nu$ is expressed in MHz and $T_e$ in K \citep{Stanislavsky2023}. 

The origin of RTE is the variation of the refractive index of the medium at radio frequencies, $\tilde{n}$, near the source of emission as
\begin{equation}\label{eq:refind}
	\tilde{n}=\sqrt{1-\frac{n_ee^2}{\pi m_e\nu^2}}.
\end{equation}
Equations (2) and (3) from \cite{Padovani2021} for the power per unit frequency emitted by an electron of energy $E$ at frequency $\nu$, $P_\nu^\mathrm{em}$, and the critical frequency, $\nu_c$, are modified as follows:
\begin{equation}
    P^\mathrm{em}_\nu(E,\vec{r}) = \frac{\sqrt3e^3}{m_ec^2}B_\perp(\vec{r})\left[1+(1-\tilde{n}^2)\left(\frac{E}{m_ec^2}\right)^2\right]^{-1/2}F(\nu_c'),
\end{equation}
\begin{equation}
    \nu_c'=\nu_c\left[1+(1-\tilde{n}^2)\left(\frac{E}{m_ec^2}\right)^2\right]^{-3/2},
\end{equation}
where $e$ and $m_e$ are the electron charge and mass, $\vec{r}$ is the position vector, $c$ is the speed of light, $\bperp$ is the magnetic-field strength on the plane of the sky, and $F$ depends on the Bessel function of order 5/3.


In Fig.~\ref{fig:ext}, we show that for ionized gas with $T_e = 10^4$ K, the fraction of recovered synchrotron emissivity, accounting for both FFA and RTE, can be as low as only a few percent of the emitted value, depending on the electron column density, $N_e$, and the average of $\bperp$ along the sight line, $\langle B_{\perp} \rangle$. 

Given a representative value of $N_{\rm H_2}$ for the protostellar-core sample in this work of $\sim10^{22}$ cm$^{-2}$ (see figure 17 in \citetalias{Pezzuto2021} and also \citealt{Bracco2017}), an ionization fraction of a few \% \citep{Fedriani2019}, and a value of $\langle B_{\perp} \rangle \approx (0.1 - 1)$ mG \citep[e.g., ][]{Girart2006}, our models show that the non-thermal radio emission at 144 MHz could be potentially extinguished, completely or partially, by the combined effects of RTE and FFA. We note that even with a reduction in column density by a couple of orders of magnitude, our conclusions would remain largely unchanged. Extinction mechanisms would still significantly impede synchrotron emission.
\subsection{Upper limits on the magnetic field in prestellar cores}
\label{ssec:prestellar}

Recent constraints on the CRe energy spectrum, $j_e$, from the Voyager spacecraft \citep{Cummings2016}, have highlighted the impact that GeV-range CRe may have on the synchrotron emission from prestellar cores in molecular clouds \citep{Dickinson2015, PadovaniGalli2018}. According to \citet{PadovaniGalli2018}, for typical magnetic field strengths in prestellar cores ($10\,\mu$G - 1 mG), meter-wavelength synchrotron emission is expected to be observable without significant attenuation below $N_{\rm H_2}$ values of approximately $6\times10^{24}$ cm$^{-2}$, which is much higher than the typical column densities of prestellar cores. 
\begin{figure}[!h]
\begin{center}
\resizebox{1\hsize}{!}{\includegraphics{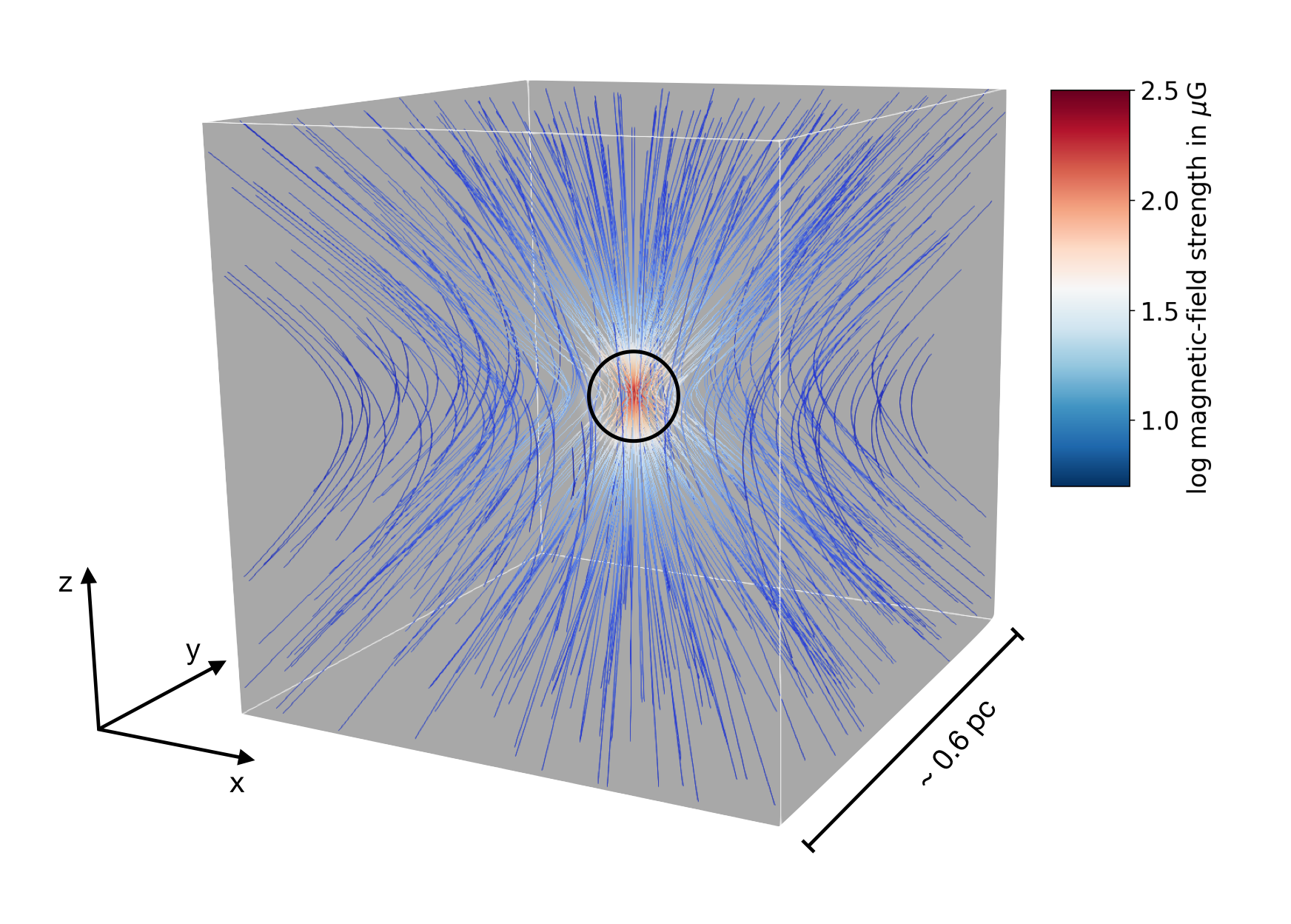}}
\caption{Representation of the magnetic-field structure and strength (in colours) of one of the prestellar-core models from \citet{Li1996}. These models are for isothermal cores at magnetostatic equilibrium. The example is shown for $\lambda_r = 1.52$. The black circle in the middle, with a diameter of $\sim$0.1~pc, indicates the region used to average the magnetic-field strength. The 3D rendering was made with the {\tt ParaView} software (\url{http://www.paraview.org/}).}
\label{fig:3Dmodel}
\end{center}
\end{figure}
In Sect.~\ref{sec:results}, we attempted to statistically detect the emission from prestellar cores using LOFAR. However, as discussed in Sect.~\ref{ssec:residu}, we did not find any significant counterparts in the corresponding residual stacked maps. Since prestellar cores are defined as starless cores, we cannot explain the lack of detection in terms of extinction of internal radio sources (e.g., jets, shocks), as we did for protostellar cores. 
\begin{figure*}[!ht]
\begin{center}
\resizebox{1.0\hsize}{!}{\includegraphics{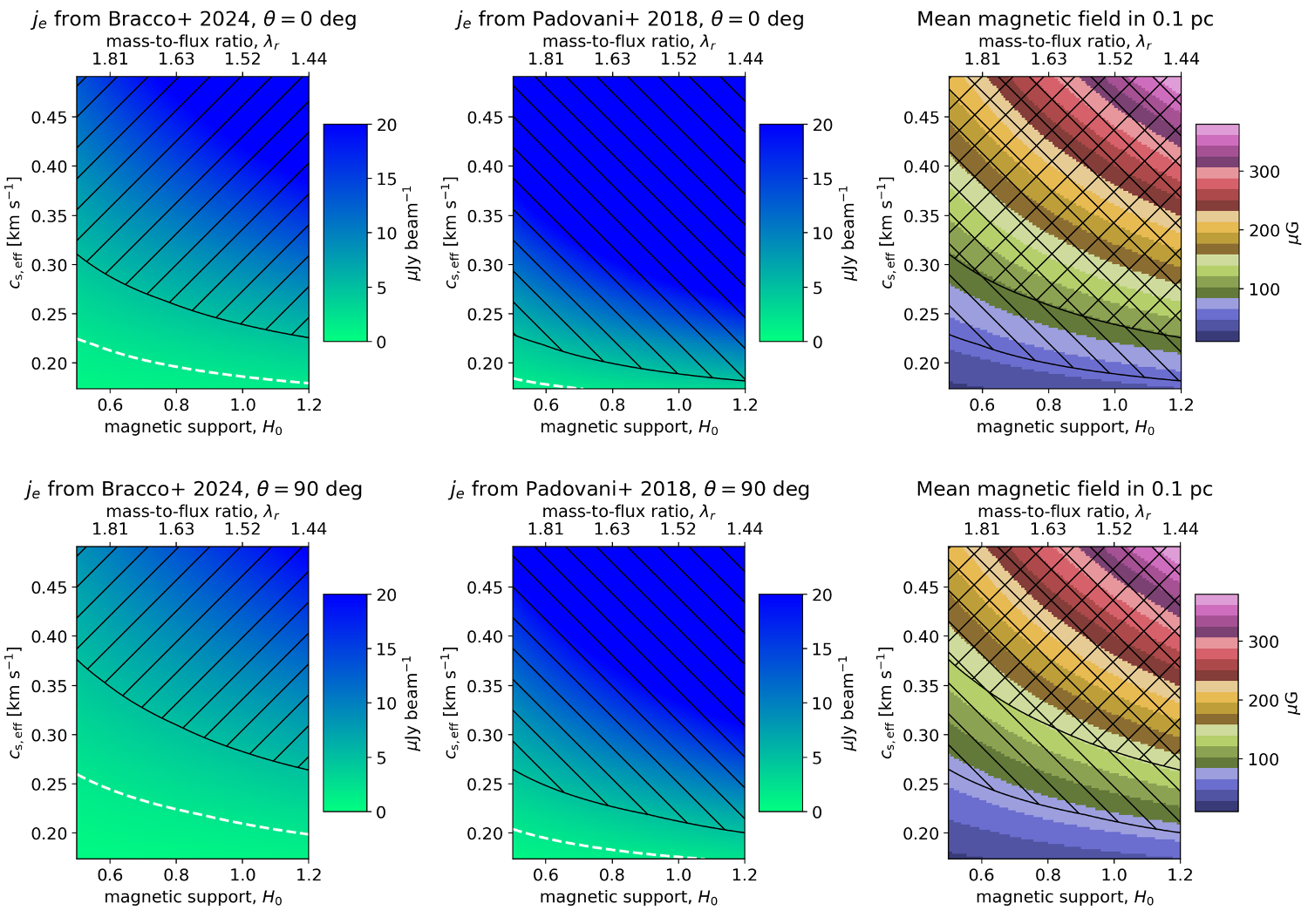}}
\caption{Estimated Stokes $I$ at 144 MHz from a grid of prestellar-core models -- see one example in Fig.~\ref{fig:3Dmodel} -- as a function of the magnetic support ($H_{\rm 0}$) and the effective sound speed in the medium ($c_{\rm s, eff}$). The corresponding mass-to-flux ratio ($\lambda_r$) is also displayed on the top of each panel. In the left and central panels two different emissivity models are considered \citep{Padovani2018,Bracco2024}. In the right panels, for each pair of parameters ($c_{\rm s, eff}$, $H_{\rm 0}$), we show the corresponding magnetic-field strength averaged within 0.1 pc from the center. In the two rows we integrate the models either edge on (top) or face on (bottom). Black hatches define the non-detection areas in the LOFAR data for Stokes $I$ values greater than 5 $\mu$Jy beam$^{-1}$. The white-dashed lines at 2 $\mu$Jy beam$^{-1}$ indicate the sensitivity in Perseus of SKA-Low for $\sim$9 hours of observations in stage AA* and $\sim$4 hours in stage AA4.}
\label{fig:model_grid}
\end{center}
\end{figure*}

In order to further investigate the missing radio emission in prestellar cores, we produced synthetic data of Stokes $I$ at 144 MHz. For the magnetic-field structure, we used the 3D axisymmetric models of starless cores of \citet{Li1996} and later extended by \citet{Galli1999}. These models, which were also employed in \citet{Padovani2011} and \citet{Padovani2013}, describe magnetostatic, isothermal, self-gravitating, axially symmetric cloud cores supported by ordered hourglass magnetic fields\footnote{Our analytical models do not account for any turbulent magnetic-field component.}. They are characterized by two main parameters: the effective sound speed, $c_{\rm s,\, eff}$, which accounts for both thermal and turbulent support to gravity; and the value of the dimensionless mass-to-flux ratio $\lambda_r$, defined as
\begin{equation}\label{eq:lambda}
    \lambda_r = 2\pi G^{1/2}\frac{M(\Phi)}{\Phi},
\end{equation}
where $G$ is the gravitational constant, and $M(\Phi)$ the mass enclosed in a sphere tangent to the flux tube $\Phi$.

Given a value of $\lambda_r$, the corresponding magnetic support to gravitational collapse is $H_0 = \phi(\pi/2)\lambda_r - 1$, where $\phi(\pi/2)$ is a dimensionless quantity. We modeled prestellar cores within boxes of approximately (0.6 pc)$^3$, which projected in 2D roughly corresponds to the size of the cut-outs used for stacking at the distance of Perseus. As an example, in Fig.~\ref{fig:3Dmodel} we show one 3D rendering of the model with $c_{\rm s,\, eff} = 0.2$ km s$^{-1}$ and $\lambda_r = 1.52$. The strength of the magnetic field varies between a few $\mu$G to more than 200 $\mu$G in the central region within 0.1 pc, as indicated by the black circle.  

Assuming a uniform spatial distribution for the CRe, we estimated $\varepsilon_{\nu}$ as before, considering two shapes of $j_e$ from \citet{Bracco2024} and \citet{Padovani2018} to quantify the uncertainties on the CRe assumptions. By convolving the synthetic synchrotron emission with the LOFAR FWHM, we estimated the average Stokes $I$ expected within 0.1 pc from the core for a grid of models defined by realistic values for prestellar cores of $c_{\rm s,\, eff} \in [0.17, 0.5]$ km s$^{-1}$ and $\lambda_r \in [1.44, 2]$ \citep[e.g.,][]{Crutcher2012}. We integrated the modeled cubes along two orthogonal orientations, $\theta$, along and across the poloidal axis of the magnetic field ($z$ and $y$ directions, respectively, in Fig.~\ref{fig:3Dmodel}).

The results of our modeling are shown in Fig.~\ref{fig:model_grid}. Regions above 5 $\mu$Jy beam$^{-1}$, or the 84-th percentile of the prestellar-core NDF (see Fig.~\ref{fig:distributions}), are indicated with black diagonal hatches. In the right column, we show the mean magnetic-field strength within 0.1 pc for each model.

We note that our statistical analysis of the LOFAR data excludes prestellar-core models with magnetic-field strengths strictly larger than 150 $\mu$G for both $j_e$ and $\theta$ cases. Nevertheless, below the LOFAR noise level, we identified a plausible range of prestellar-core models that could be detected at higher sensitivity with ordered magnetic-field strengths of $100\, \mu$G, or less, and with values of $c_{\rm s,\, eff}$ close to 0.2 km s$^{-1}$, which corresponds to the sound speed of isothermal cores with a temperature of 10 K. 

Despite the simplicity of our model, we reached constraints that are compatible with the very few observational evidences of magnetic-field strengths in prestellar cores, ranging from tens to hundreds of $\mu$G \citep[e.g., ][ and references therein]{Pattle2023}. We also notice that most of these observational estimates only represent upper limits to the magnetic-field strength as they rely on the standard Davis-Chandrasekhar-Fermi method applied to interstellar dust polarization measurements, which can statistically overestimate the interstellar magnetic-field strength up to an order of magnitude \citep[e.g., ][]{Skalidis2021}.  

Prestellar cores with magnetic-field strengths below 100 $\mu$G will be  observable, and likely detected, with SKA-Low. We used the SKA-Low sensitivity calculator and time estimator\footnote{\url{http://sensitivity-calculator.skao.int/low}} \citep{Sokolowski2022}, to investigate how many hours would be needed to reach a sensitivity of 2 $\mu$Jy beam$^{-1}$ in the residual stacking of the sample of 348 prestellar cores in Perseus at 144 MHz under observational conditions similar to those in LoTSS. In the left and central panels of Fig.~\ref{fig:model_grid}, we show the 2 $\mu$Jy beam$^{-1}$ level with a white-dashed line, which should be attainable in only 9 hours and 4 hours of observing time, during the preliminary stage of array assembly\footnote{For more details on the array assembly key information visit, \url{http://www.skao.int/en/science-users/159/scientific-timeline}}, AA*, and the final stage AA4, respectively.

\section{Summary and conclusion}\label{sec:summary}

Cosmic rays (CRs) are considered an essential ingredient in the star formation process, determining the ionization rate of molecular gas and its interaction with magnetic fields \citep[e.g.,][]{Padovani2020}. The electronic GeV component (CRe) is expected to produce mostly non-thermal synchrotron radiation when interacting with the interstellar magnetic field. Moreover, star formation, protostellar activity, and jets are also thought to locally accelerate CRe in shocks \citep{Padovani2015, Padovani2016, Gaches2018, Padovani2021b}. Despite these theoretical expectations, however, non-thermal emission from molecular clouds in the Milky Way has been observed in only a few cases, such as diffuse emission in the Orion-Taurus ridge \citep{Bracco2023} and compact regions around one protostellar jet in DG Tau A \citep{FeeneyJohansson2019}.

While non-thermal radiation in the GHz range can be strongly contaminated by thermal free-free emission \citep[e.g.,][]{Tobin2016}, the low-frequency range in the radio represents the most promising observational window to detect synchrotron emission from molecular clouds \citep[e.g.,][]{Dickinson2015}. Although current facilities do not easily achieve the required sensitivity to detect emission from single objects \citep{PadovaniGalli2018}, in this work, we have conducted the first statistical analysis aimed at extracting non-thermal radio emission from hundreds of stellar embryos, both prestellar and protostellar cores, by stacking LOFAR data at 144 MHz from LoTSS \citep{Shimwell2022} toward the Perseus molecular cloud.

Among the 353 prestellar cores and 132 protostellar cores identified in Perseus by \citetalias{Pezzuto2021} using {\it Herschel} data, we found 5 prestellar cores and 18 protostellar cores with significant excess emission in LOFAR data. However, for all these cores, we estimated that the correspondence between far-infrared and radio emissions is likely due to extragalactic contaminations in the {\it Herschel} catalog by \citetalias{Pezzuto2021} (see Sect.~\ref{ssec:stack1} and Sect.~\ref{ssec:sources}). 
By stacking the remaining prestellar and protostellar cores separately, we could not detect any significant excess in LOFAR at the levels of 5 and 8 $\mu$Jy beam$^{-1}$ for prestellar cores and protostellar cores, respectively (see Sect.~\ref{ssec:residu}). We interpreted these results in two ways: 
in the case of protostellar cores, we propose that the radio emission could be completely extinguished by either free-free absorption or the Razin-Tsytovich effect along the sight line (see Sect.~\ref{ssec:ext}); 
in the case of prestellar cores, we concluded that, even after stacking, we were primarily limited by the sensitivity of LoTSS. We compared the observational results with synthetic data from the 3D analytical models of magnetostatic-isothermal cores by \citet{Li1996} and set upper limits on the ordered magnetic-field strength that prestellar cores should have not to be detected in the stacking of LOFAR data. Statistically, prestellar cores in Perseus should have magnetic-field strengths below 100 $\mu$G within 0.1 pc from the core center (see Sect.~\ref{ssec:prestellar}). 
Using the SKA sensitivity calculator, we predicted that such prestellar cores in Perseus will be statistically detectable by SKA-Low in 9 and 4 hours observing time, using the array assemblies AA* and AA4, respectively.

\begin{acknowledgements} 
We acknowledge the referee's thorough and insightful comments, which have significantly improved the clarity and content of this manuscript. We are grateful to Tim Shimwell, Florent Mertens, Philippe Zarka, Giulia Macario, Francesca Bacciotti and Marta Fatovi\'c for useful discussions. We thank Philippe Andr\'e for providing us with the column-density map of the Perseus cloud. AB and DG acknowledge financial support from the INAF initiative ``IAF Astronomy Fellowships in Italy'' (grant name MEGASKAT) and the INAF minigrant PACIFISM, respectively. We also acknowledge the inspiration provided by Delphine and Elio. AD acknowledges support by the BMBF Verbundforschung under the grant 05A20STA. The J{\"u}lich LOFAR Long Term Archive and the German LOFAR network are both coordinated and operated by the J{\"u}lich Supercomputing Centre (JSC), and computing resources on the supercomputer JUWELS at JSC were provided by the Gauss Centre for Supercomputing e.V. (grant CHTB00) through the John von Neumann Institute for Computing (NIC).
LOFAR data products were provided by the LOFAR Surveys Key Science project
(LSKSP; https://lofar-surveys.org/) and were derived from observations with the International LOFAR Telescope (ILT). LOFAR (van Haarlem et al. 2013) is the Low Frequency Array designed and constructed by ASTRON. It has observing, data processing, and data storage facilities in several countries, that are owned by various parties (each with their own funding
sources), and that are collectively operated by the ILT foundation under a joint scientific policy. The efforts of the LSKSP have benefited from funding from the European Research Council, NOVA, NWO, CNRS-INSU, the SURF Co-operative, the UK Science and Technology Funding Council and the J\"ulich Supercomputing Centre.
In the analysis we made use of {\tt astropy} \citep{astropy2018}, {\tt scipy} \citep{Virtanen2020}, and {\tt numpy} \citep{Harris2020}. This research has made use of “Aladin sky atlas” developed at CDS, Strasbourg
Observatory, France.
\end{acknowledgements}

\bibliographystyle{aa}
\bibliography{aanda.bbl}

\begin{thebibliography}{85}
\expandafter\ifx\csname natexlab\endcsname\relax\def\natexlab#1{#1}\fi

\bibitem[{{Andre} {et~al.}(2000){Andre}, {Ward-Thompson}, \& {Barsony}}]{Andre2000}
{Andre}, P., {Ward-Thompson}, D., \& {Barsony}, M. 2000, in Protostars and Planets IV, ed. V.~{Mannings}, A.~P. {Boss}, \& S.~S. {Russell}, 59

\bibitem[{{Anglada}(1995)}]{Anglada1995}
{Anglada}, G. 1995, in Revista Mexicana de Astronomia y Astrofisica Conference Series, Vol.~1, Revista Mexicana de Astronomia y Astrofisica Conference Series, ed. S.~{Lizano} \& J.~M. {Torrelles}, 67

\bibitem[{{Astropy Collaboration} {et~al.}(2018){Astropy Collaboration}, {Price-Whelan}, {Sip{\H{o}}cz}, {G{\"u}nther}, {Lim}, {Crawford}, {Conseil}, {Shupe}, {Craig}, {Dencheva}, {Ginsburg}, {VanderPlas}, {Bradley}, {P{\'e}rez-Su{\'a}rez}, {de Val-Borro}, {Aldcroft}, {Cruz}, {Robitaille}, {Tollerud}, {Ardelean}, {Babej}, {Bach}, {Bachetti}, {Bakanov}, {Bamford}, {Barentsen}, {Barmby}, {Baumbach}, {Berry}, {Biscani}, {Boquien}, {Bostroem}, {Bouma}, {Brammer}, {Bray}, {Breytenbach}, {Buddelmeijer}, {Burke}, {Calderone}, {Cano Rodr{\'\i}guez}, {Cara}, {Cardoso}, {Cheedella}, {Copin}, {Corrales}, {Crichton}, {D'Avella}, {Deil}, {Depagne}, {Dietrich}, {Donath}, {Droettboom}, {Earl}, {Erben}, {Fabbro}, {Ferreira}, {Finethy}, {Fox}, {Garrison}, {Gibbons}, {Goldstein}, {Gommers}, {Greco}, {Greenfield}, {Groener}, {Grollier}, {Hagen}, {Hirst}, {Homeier}, {Horton}, {Hosseinzadeh}, {Hu}, {Hunkeler}, {Ivezi{\'c}}, {Jain}, {Jenness}, {Kanarek}, {Kendrew}, {Kern}, {Kerzendorf}, {Khvalko}, {King}, {Kirkby}, {Kulkarni},
  {Kumar}, {Lee}, {Lenz}, {Littlefair}, {Ma}, {Macleod}, {Mastropietro}, {McCully}, {Montagnac}, {Morris}, {Mueller}, {Mumford}, {Muna}, {Murphy}, {Nelson}, {Nguyen}, {Ninan}, {N{\"o}the}, {Ogaz}, {Oh}, {Parejko}, {Parley}, {Pascual}, {Patil}, {Patil}, {Plunkett}, {Prochaska}, {Rastogi}, {Reddy Janga}, {Sabater}, {Sakurikar}, {Seifert}, {Sherbert}, {Sherwood-Taylor}, {Shih}, {Sick}, {Silbiger}, {Singanamalla}, {Singer}, {Sladen}, {Sooley}, {Sornarajah}, {Streicher}, {Teuben}, {Thomas}, {Tremblay}, {Turner}, {Terr{\'o}n}, {van Kerkwijk}, {de la Vega}, {Watkins}, {Weaver}, {Whitmore}, {Woillez}, {Zabalza}, \& {Astropy Contributors}}]{astropy2018}
{Astropy Collaboration}, {Price-Whelan}, A.~M., {Sip{\H{o}}cz}, B.~M., {et~al.} 2018, \aj, 156, 123

\bibitem[{{Ballesteros-Paredes} {et~al.}(2020){Ballesteros-Paredes}, {Andr{\'e}}, {Hennebelle}, {Klessen}, {Kruijssen}, {Chevance}, {Nakamura}, {Adamo}, \& {V{\'a}zquez-Semadeni}}]{BallesterosParedes2020}
{Ballesteros-Paredes}, J., {Andr{\'e}}, P., {Hennebelle}, P., {et~al.} 2020, \ssr, 216, 76

\bibitem[{{Boch} \& {Fernique}(2014)}]{Boch2014}
{Boch}, T. \& {Fernique}, P. 2014, in Astronomical Society of the Pacific Conference Series, Vol. 485, Astronomical Data Analysis Software and Systems XXIII, ed. N.~{Manset} \& P.~{Forshay}, 277

\bibitem[{{Bracco} {et~al.}(2020){Bracco}, {Bresnahan}, {Palmeirim}, {Arzoumanian}, {Andr{\'e}}, {Ward-Thompson}, \& {Marchal}}]{Bracco2020c}
{Bracco}, A., {Bresnahan}, D., {Palmeirim}, P., {et~al.} 2020, \aap, 644, A5

\bibitem[{{Bracco} {et~al.}(2024){Bracco}, {Padovani}, \& {Galli}}]{Bracco2024}
{Bracco}, A., {Padovani}, M., \& {Galli}, D. 2024, \aap, 686, A52

\bibitem[{{Bracco} {et~al.}(2023){Bracco}, {Padovani}, \& {Soler}}]{Bracco2023}
{Bracco}, A., {Padovani}, M., \& {Soler}, J.~D. 2023, \aap, 677, L11

\bibitem[{{Bracco} {et~al.}(2017){Bracco}, {Palmeirim}, {Andr{\'e}}, {Adam}, {Ade}, {Bacmann}, {Beelen}, {Beno{\^\i}t}, {Bideaud}, {Billot}, {Bourrion}, {Calvo}, {Catalano}, {Coiffard}, {Comis}, {D'Addabbo}, {D{\'e}sert}, {Didelon}, {Doyle}, {Goupy}, {K{\"o}nyves}, {Kramer}, {Lagache}, {Leclercq}, {Mac{\'\i}as-P{\'e}rez}, {Maury}, {Mauskopf}, {Mayet}, {Monfardini}, {Motte}, {Pajot}, {Pascale}, {Peretto}, {Perotto}, {Pisano}, {Ponthieu}, {Rev{\'e}ret}, {Rigby}, {Ritacco}, {Rodriguez}, {Romero}, {Roy}, {Ruppin}, {Schuster}, {Sievers}, {Triqueneaux}, {Tucker}, \& {Zylka}}]{Bracco2017}
{Bracco}, A., {Palmeirim}, P., {Andr{\'e}}, P., {et~al.} 2017, \aap, 604, A52

\bibitem[{{Chen} {et~al.}(2022){Chen}, {Ebert}, {Mazzarella}, {Frayer}, {Terek}, {Chan}, {Cook}, {Lo}, {Schmitz}, \& {Wu}}]{Chen2022}
{Chen}, T.~X., {Ebert}, R., {Mazzarella}, J.~M., {et~al.} 2022, \pasp, 134, 014501

\bibitem[{{Condon} {et~al.}(1998){Condon}, {Cotton}, {Greisen}, {Yin}, {Perley}, {Taylor}, \& {Broderick}}]{Condon1998}
{Condon}, J.~J., {Cotton}, W.~D., {Greisen}, E.~W., {et~al.} 1998, \aj, 115, 1693

\bibitem[{{Coud{\'e}} {et~al.}(2019){Coud{\'e}}, {Bastien}, {Houde}, {Sadavoy}, {Friesen}, {Di Francesco}, {Johnstone}, {Mairs}, {Hasegawa}, {Kwon}, {Lai}, {Qiu}, {Ward-Thompson}, {Berry}, {Chen}, {Fiege}, {Franzmann}, {Hatchell}, {Lacaille}, {Matthews}, {Moriarty-Schieven}, {Pon}, {Andr{\'e}}, {Arzoumanian}, {Aso}, {Byun}, {Eswaraiah}, {Chen}, {Chen}, {Ching}, {Cho}, {Choi}, {Chrysostomou}, {Chung}, {Doi}, {Drabek-Maunder}, {Dowell}, {Eyres}, {Falle}, {Friberg}, {Fuller}, {Furuya}, {Gledhill}, {Graves}, {Greaves}, {Griffin}, {Gu}, {Hayashi}, {Hoang}, {Holland}, {Inoue}, {Inutsuka}, {Iwasaki}, {Jeong}, {Kanamori}, {Kataoka}, {Kang}, {Kang}, {Kang}, {Kawabata}, {Kemper}, {Kim}, {Kim}, {Kim}, {Kim}, {Kim}, {Kim}, {Kirk}, {Kobayashi}, {Koch}, {Kwon}, {Lee}, {Lee}, {Lee}, {Li}, {Li}, {Li}, {Liu}, {Liu}, {Liu}, {Liu}, {van Loo}, {Lyo}, {Matsumura}, {Nagata}, {Nakamura}, {Nakanishi}, {Ohashi}, {Onaka}, {Parsons}, {Pattle}, {Peretto}, {Pyo}, {Qian}, {Rao}, {Rawlings}, {Retter}, {Richer}, {Rigby}, {Robitaille},
  {Saito}, {Savini}, {Scaife}, {Seta}, {Shinnaga}, {Soam}, {Tamura}, {Tang}, {Tomisaka}, {Tsukamoto}, {Wang}, {Wang}, {Whitworth}, {Yen}, {Yoo}, {Yuan}, {Zenko}, {Zhang}, {Zhang}, {Zhou}, {Zhu}, \& {B-fields In STar-forming Regions Observations (BISTRO Collaboration}}]{Coude2019}
{Coud{\'e}}, S., {Bastien}, P., {Houde}, M., {et~al.} 2019, \apj, 877, 88

\bibitem[{{Coughlan} {et~al.}(2017){Coughlan}, {Ainsworth}, {Eisl{\"o}ffel}, {Hoeft}, {Drabent}, {Scaife}, {Ray}, {Bell}, {Broderick}, {Corbel}, {Grie{\ss}meier}, {van der Horst}, {van Leeuwen}, {Markoff}, {Pietka}, {Stewart}, {Wijers}, \& {Zarka}}]{Coughlan2017}
{Coughlan}, C.~P., {Ainsworth}, R.~E., {Eisl{\"o}ffel}, J., {et~al.} 2017, \apj, 834, 206

\bibitem[{{Crutcher}(2012)}]{Crutcher2012}
{Crutcher}, R.~M. 2012, \araa, 50, 29

\bibitem[{{Crutcher} {et~al.}(2010){Crutcher}, {Wandelt}, {Heiles}, {Falgarone}, \& {Troland}}]{Crutcher2010}
{Crutcher}, R.~M., {Wandelt}, B., {Heiles}, C., {Falgarone}, E., \& {Troland}, T.~H. 2010, \apj, 725, 466

\bibitem[{{Cummings} {et~al.}(2016){Cummings}, {Stone}, {Heikkila}, {Lal}, {Webber}, {J{\'o}hannesson}, {Moskalenko}, {Orlando}, \& {Porter}}]{Cummings2016}
{Cummings}, A.~C., {Stone}, E.~C., {Heikkila}, B.~C., {et~al.} 2016, \apj, 831, 18

\bibitem[{{de Gasperin} {et~al.}(2019){de Gasperin}, {Dijkema}, {Drabent}, {Mevius}, {Rafferty}, {van Weeren}, {Br{\"u}ggen}, {Callingham}, {Emig}, {Heald}, {Intema}, {Morabito}, {Offringa}, {Oonk}, {Orr{\`u}}, {R{\"o}ttgering}, {Sabater}, {Shimwell}, {Shulevski}, \& {Williams}}]{deGasperin2019}
{de Gasperin}, F., {Dijkema}, T.~J., {Drabent}, A., {et~al.} 2019, \aap, 622, A5

\bibitem[{{Dewdney} {et~al.}(2009){Dewdney}, {Hall}, {Schilizzi}, \& {Lazio}}]{Dewdney2009}
{Dewdney}, P.~E., {Hall}, P.~J., {Schilizzi}, R.~T., \& {Lazio}, T.~J.~L.~W. 2009, IEEE Proceedings, 97, 1482

\bibitem[{{Dickinson} {et~al.}(2015){Dickinson}, {Beck}, {Crocker}, {Crutcher}, {Davies}, {Ferri{\`e}re}, {Fuller}, {Jaffe}, {Jones}, {Leahy}, {Murphy}, {Peel}, {Orlando}, {Porter}, {Protheroe}, {Strong}, {Robishaw}, {Watson}, \& {Yusef-Zadeh}}]{Dickinson2015}
{Dickinson}, C., {Beck}, R., {Crocker}, R., {et~al.} 2015, in Advancing Astrophysics with the Square Kilometre Array (AASKA14), 102

\bibitem[{{Doi} {et~al.}(2021){Doi}, {Hasegawa}, {Bastien}, {Tahani}, {Arzoumanian}, {Coud{\'e}}, {Matsumura}, {Sadavoy}, {Hull}, {Shimajiri}, {Furuya}, {Johnstone}, {Plume}, {Inutsuka}, {Kwon}, \& {Tamura}}]{Doi2021}
{Doi}, Y., {Hasegawa}, T., {Bastien}, P., {et~al.} 2021, \apj, 914, 122

\bibitem[{{Drabent} {et~al.}(2019){Drabent}, {Hoeft}, {Mechev}, {Oonk}, {Shimwell}, {Sweijen}, {Danezi}, {Schrijvers}, {Manzano}, {Tsigenov}, {Dettmar}, {Br{\"u}ggen}, \& {Schwarz}}]{Drabent2019}
{Drabent}, A., {Hoeft}, M., {Mechev}, A.~P., {et~al.} 2019, arXiv e-prints, arXiv:1910.13835

\bibitem[{{Dunham} {et~al.}(2008){Dunham}, {Crapsi}, {Evans}, {Bourke}, {Huard}, {Myers}, \& {Kauffmann}}]{Dunham2008}
{Dunham}, M.~M., {Crapsi}, A., {Evans}, Neal~J., I., {et~al.} 2008, \apjs, 179, 249

\bibitem[{{Evans} {et~al.}(2009){Evans}, {Dunham}, {J{\o}rgensen}, {Enoch}, {Mer{\'\i}n}, {van Dishoeck}, {Alcal{\'a}}, {Myers}, {Stapelfeldt}, {Huard}, {Allen}, {Harvey}, {van Kempen}, {Blake}, {Koerner}, {Mundy}, {Padgett}, \& {Sargent}}]{Evans2009}
{Evans}, Neal~J., I., {Dunham}, M.~M., {J{\o}rgensen}, J.~K., {et~al.} 2009, \apjs, 181, 321

\bibitem[{{Fedriani} {et~al.}(2019){Fedriani}, {Caratti o Garatti}, {Purser}, {Sanna}, {Tan}, {Garcia-Lopez}, {Ray}, {Coffey}, {Stecklum}, \& {Hoare}}]{Fedriani2019}
{Fedriani}, R., {Caratti o Garatti}, A., {Purser}, S.~J.~D., {et~al.} 2019, Nature Communications, 10, 3630

\bibitem[{{Feeney-Johansson} {et~al.}(2019){Feeney-Johansson}, {Purser}, {Ray}, {Eisl{\"o}ffel}, {Hoeft}, {Drabent}, \& {Ainsworth}}]{FeeneyJohansson2019}
{Feeney-Johansson}, A., {Purser}, S. J.~D., {Ray}, T.~P., {et~al.} 2019, \apjl, 885, L7

\bibitem[{{Gabici}(2022)}]{Gabici2022}
{Gabici}, S. 2022, \aapr, 30, 4

\bibitem[{{Gaches} \& {Offner}(2018)}]{Gaches2018}
{Gaches}, B. A.~L. \& {Offner}, S. S.~R. 2018, \apj, 861, 87

\bibitem[{{Galli} {et~al.}(1999){Galli}, {Lizano}, {Li}, {Adams}, \& {Shu}}]{Galli1999}
{Galli}, D., {Lizano}, S., {Li}, Z.~Y., {Adams}, F.~C., \& {Shu}, F.~H. 1999, \apj, 521, 630

\bibitem[{{Ginzburg} \& {Syrovatskii}(1965)}]{Ginzburg1965}
{Ginzburg}, V.~L. \& {Syrovatskii}, S.~I. 1965, \araa, 3, 297

\bibitem[{{Girart} {et~al.}(2006){Girart}, {Rao}, \& {Marrone}}]{Girart2006}
{Girart}, J.~M., {Rao}, R., \& {Marrone}, D.~P. 2006, Science, 313, 812

\bibitem[{{Grenier} {et~al.}(2015){Grenier}, {Black}, \& {Strong}}]{Grenier2015}
{Grenier}, I.~A., {Black}, J.~H., \& {Strong}, A.~W. 2015, \araa, 53, 199

\bibitem[{{Hacar} {et~al.}(2023){Hacar}, {Clark}, {Heitsch}, {Kainulainen}, {Panopoulou}, {Seifried}, \& {Smith}}]{Hacar2023}
{Hacar}, A., {Clark}, S.~E., {Heitsch}, F., {et~al.} 2023, in Astronomical Society of the Pacific Conference Series, Vol. 534, Protostars and Planets VII, ed. S.~{Inutsuka}, Y.~{Aikawa}, T.~{Muto}, K.~{Tomida}, \& M.~{Tamura}, 153

\bibitem[{{Harris} {et~al.}(2020){Harris}, {Millman}, {van der Walt}, {Gommers}, {Virtanen}, {Cournapeau}, {Wieser}, {Taylor}, {Berg}, {Smith}, {Kern}, {Picus}, {Hoyer}, {van Kerkwijk}, {Brett}, {Haldane}, {del R{\'\i}o}, {Wiebe}, {Peterson}, {G{\'e}rard-Marchant}, {Sheppard}, {Reddy}, {Weckesser}, {Abbasi}, {Gohlke}, \& {Oliphant}}]{Harris2020}
{Harris}, C.~R., {Millman}, K.~J., {van der Walt}, S.~J., {et~al.} 2020, \nat, 585, 357

\bibitem[{{Heiles} \& {Troland}(2003)}]{Heiles2003}
{Heiles}, C. \& {Troland}, T.~H. 2003, \apj, 586, 1067

\bibitem[{{Hennebelle} \& {Inutsuka}(2019)}]{Hennebelle2019}
{Hennebelle}, P. \& {Inutsuka}, S.-i. 2019, Frontiers in Astronomy and Space Sciences, 6, 5

\bibitem[{{Intema} {et~al.}(2017){Intema}, {Jagannathan}, {Mooley}, \& {Frail}}]{Intema2017}
{Intema}, H.~T., {Jagannathan}, P., {Mooley}, K.~P., \& {Frail}, D.~A. 2017, \aap, 598, A78

\bibitem[{{Ivlev} {et~al.}(2021){Ivlev}, {Silsbee}, {Padovani}, \& {Galli}}]{Ivlev2021}
{Ivlev}, A.~V., {Silsbee}, K., {Padovani}, M., \& {Galli}, D. 2021, \apj, 909, 107

\bibitem[{{Lee} {et~al.}(2010){Lee}, {Le Floc'h}, {Sanders}, {Frayer}, {Arnouts}, {Ilbert}, {Aussel}, {Salvato}, {Scoville}, \& {Kartaltepe}}]{Lee2010}
{Lee}, N., {Le Floc'h}, E., {Sanders}, D.~B., {et~al.} 2010, \apj, 717, 175

\bibitem[{{Li} \& {Shu}(1996)}]{Li1996}
{Li}, Z.-Y. \& {Shu}, F.~H. 1996, \apj, 472, 211

\bibitem[{{Magnier} {et~al.}(2020){Magnier}, {Schlafly}, {Finkbeiner}, {Tonry}, {Goldman}, {R{\"o}ser}, {Schilbach}, {Casertano}, {Chambers}, {Flewelling}, {Huber}, {Price}, {Sweeney}, {Waters}, {Denneau}, {Draper}, {Hodapp}, {Jedicke}, {Kaiser}, {Kudritzki}, {Metcalfe}, {Stubbs}, \& {Wainscoat}}]{Magnier2020}
{Magnier}, E.~A., {Schlafly}, E.~F., {Finkbeiner}, D.~P., {et~al.} 2020, \apjs, 251, 6

\bibitem[{{Maury} {et~al.}(2022){Maury}, {Hennebelle}, \& {Girart}}]{Maury2022}
{Maury}, A., {Hennebelle}, P., \& {Girart}, J.~M. 2022, Frontiers in Astronomy and Space Sciences, 9, 949223

\bibitem[{{Momferratos} {et~al.}(2014){Momferratos}, {Lesaffre}, {Falgarone}, \& {Pineau des For{\^e}ts}}]{Momferratos2014}
{Momferratos}, G., {Lesaffre}, P., {Falgarone}, E., \& {Pineau des For{\^e}ts}, G. 2014, \mnras, 443, 86

\bibitem[{{Offringa} {et~al.}(2012){Offringa}, {van de Gronde}, \& {Roerdink}}]{Offringa2012}
{Offringa}, A.~R., {van de Gronde}, J.~J., \& {Roerdink}, J.~B.~T.~M. 2012, \aap, 539, A95

\bibitem[{{Padovani} {et~al.}(2022){Padovani}, {Bialy}, {Galli}, {Ivlev}, {Grassi}, {Scarlett}, {Rehill}, {Zammit}, {Fursa}, \& {Bray}}]{Padovani2022}
{Padovani}, M., {Bialy}, S., {Galli}, D., {et~al.} 2022, \aap, 658, A189

\bibitem[{{Padovani} {et~al.}(2021{\natexlab{a}}){Padovani}, {Bracco}, {Jeli{\'c}}, {Galli}, \& {Bellomi}}]{Padovani2021}
{Padovani}, M., {Bracco}, A., {Jeli{\'c}}, V., {Galli}, D., \& {Bellomi}, E. 2021{\natexlab{a}}, \aap, 651, A116

\bibitem[{{Padovani} \& {Galli}(2011)}]{Padovani2011}
{Padovani}, M. \& {Galli}, D. 2011, \aap, 530, A109

\bibitem[{{Padovani} \& {Galli}(2018)}]{PadovaniGalli2018}
{Padovani}, M. \& {Galli}, D. 2018, \aap, 620, L4

\bibitem[{{Padovani} {et~al.}(2018){Padovani}, {Galli}, {Ivlev}, {Caselli}, \& {Ferrara}}]{Padovani2018}
{Padovani}, M., {Galli}, D., {Ivlev}, A.~V., {Caselli}, P., \& {Ferrara}, A. 2018, \aap, 619, A144

\bibitem[{{Padovani} {et~al.}(2013){Padovani}, {Hennebelle}, \& {Galli}}]{Padovani2013}
{Padovani}, M., {Hennebelle}, P., \& {Galli}, D. 2013, \aap, 560, A114

\bibitem[{{Padovani} {et~al.}(2015){Padovani}, {Hennebelle}, {Marcowith}, \& {Ferri{\`e}re}}]{Padovani2015}
{Padovani}, M., {Hennebelle}, P., {Marcowith}, A., \& {Ferri{\`e}re}, K. 2015, \aap, 582, L13

\bibitem[{{Padovani} {et~al.}(2020){Padovani}, {Ivlev}, {Galli}, {Offner}, {Indriolo}, {Rodgers-Lee}, {Marcowith}, {Girichidis}, {Bykov}, \& {Kruijssen}}]{Padovani2020}
{Padovani}, M., {Ivlev}, A.~V., {Galli}, D., {et~al.} 2020, \ssr, 216, 29

\bibitem[{{Padovani} {et~al.}(2021{\natexlab{b}}){Padovani}, {Marcowith}, {Galli}, {Hunt}, \& {Fontani}}]{Padovani2021b}
{Padovani}, M., {Marcowith}, A., {Galli}, D., {Hunt}, L.~K., \& {Fontani}, F. 2021{\natexlab{b}}, \aap, 649, A149

\bibitem[{{Padovani} {et~al.}(2016){Padovani}, {Marcowith}, {Hennebelle}, \& {Ferri{\`e}re}}]{Padovani2016}
{Padovani}, M., {Marcowith}, A., {Hennebelle}, P., \& {Ferri{\`e}re}, K. 2016, \aap, 590, A8

\bibitem[{{Pattle} {et~al.}(2023){Pattle}, {Fissel}, {Tahani}, {Liu}, \& {Ntormousi}}]{Pattle2023}
{Pattle}, K., {Fissel}, L., {Tahani}, M., {Liu}, T., \& {Ntormousi}, E. 2023, in Astronomical Society of the Pacific Conference Series, Vol. 534, Protostars and Planets VII, ed. S.~{Inutsuka}, Y.~{Aikawa}, T.~{Muto}, K.~{Tomida}, \& M.~{Tamura}, 193

\bibitem[{{Pezzuto} {et~al.}(2021){Pezzuto}, {Benedettini}, {Di Francesco}, {Palmeirim}, {Sadavoy}, {Schisano}, {Li Causi}, {Andr{\'e}}, {Arzoumanian}, {Bernard}, {Bontemps}, {Elia}, {Fiorellino}, {Kirk}, {K{\"o}nyves}, {Ladjelate}, {Men'shchikov}, {Motte}, {Piccotti}, {Schneider}, {Spinoglio}, {Ward-Thompson}, \& {Wilson}}]{Pezzuto2021}
{Pezzuto}, S., {Benedettini}, M., {Di Francesco}, J., {et~al.} 2021, \aap, 645, A55 (P21)

\bibitem[{{Pilbratt} {et~al.}(2010){Pilbratt}, {Riedinger}, {Passvogel}, {Crone}, {Doyle}, {Gageur}, {Heras}, {Jewell}, {Metcalfe}, {Ott}, \& {Schmidt}}]{Pilbratt2010}
{Pilbratt}, G.~L., {Riedinger}, J.~R., {Passvogel}, T., {et~al.} 2010, \aap, 518, L1

\bibitem[{{Pinto} {et~al.}(2008){Pinto}, {Galli}, \& {Bacciotti}}]{Pinto2008}
{Pinto}, C., {Galli}, D., \& {Bacciotti}, F. 2008, \aap, 484, 1

\bibitem[{{Planck Collaboration results. I.}(2016)}]{PlanckI2016}
{Planck Collaboration results. I.} 2016, \aap, 594, A1

\bibitem[{{Planck int. results. XXXII.}(2016)}]{PlanckXXXII}
{Planck int. results. XXXII.} 2016, \aap, 586, A135

\bibitem[{{Planck int. results. XXXV.}(2016)}]{PlanckXXXV}
{Planck int. results. XXXV.} 2016, \aap, 586, A138

\bibitem[{{Purser} {et~al.}(2018){Purser}, {Ainsworth}, {Ray}, {Green}, {Taylor}, \& {Scaife}}]{Purser2018}
{Purser}, S.~J.~D., {Ainsworth}, R.~E., {Ray}, T.~P., {et~al.} 2018, \mnras, 481, 5532

\bibitem[{{Rodr{\'\i}guez-Kamenetzky} {et~al.}(2017){Rodr{\'\i}guez-Kamenetzky}, {Carrasco-Gonz{\'a}lez}, {Araudo}, {Romero}, {Torrelles}, {Rodr{\'\i}guez}, {Anglada}, {Mart{\'\i}}, {Perucho}, \& {Valotto}}]{Rodriguez2017}
{Rodr{\'\i}guez-Kamenetzky}, A., {Carrasco-Gonz{\'a}lez}, C., {Araudo}, A., {et~al.} 2017, \apj, 851, 16

\bibitem[{{Rodr{\'\i}guez-Kamenetzky} {et~al.}(2016){Rodr{\'\i}guez-Kamenetzky}, {Carrasco-Gonz{\'a}lez}, {Araudo}, {Torrelles}, {Anglada}, {Mart{\'\i}}, {Rodr{\'\i}guez}, \& {Valotto}}]{Rodriguez2016}
{Rodr{\'\i}guez-Kamenetzky}, A., {Carrasco-Gonz{\'a}lez}, C., {Araudo}, A., {et~al.} 2016, \apj, 818, 27

\bibitem[{{Sanna} {et~al.}(2019){Sanna}, {Moscadelli}, {Goddi}, {Beltr{\'a}n}, {Brogan}, {Caratti o Garatti}, {Carrasco-Gonz{\'a}lez}, {Hunter}, {Massi}, \& {Padovani}}]{Sanna2019}
{Sanna}, A., {Moscadelli}, L., {Goddi}, C., {et~al.} 2019, \aap, 623, L3

\bibitem[{{Scaife} \& {Heald}(2012)}]{Scaife2012}
{Scaife}, A. M.~M. \& {Heald}, G.~H. 2012, \mnras, 423, L30

\bibitem[{{Shimwell} {et~al.}(2022){Shimwell}, {Hardcastle}, {Tasse}, {Best}, {R{\"o}ttgering}, {Williams}, {Botteon}, {Drabent}, {Mechev}, {Shulevski}, {van Weeren}, {Bester}, {Br{\"u}ggen}, {Brunetti}, {Callingham}, {Chy{\.z}y}, {Conway}, {Dijkema}, {Duncan}, {de Gasperin}, {Hale}, {Haverkorn}, {Hugo}, {Jackson}, {Mevius}, {Miley}, {Morabito}, {Morganti}, {Offringa}, {Oonk}, {Rafferty}, {Sabater}, {Smith}, {Schwarz}, {Smirnov}, {O'Sullivan}, {Vedantham}, {White}, {Albert}, {Alegre}, {Asabere}, {Bacon}, {Bonafede}, {Bonnassieux}, {Brienza}, {Bilicki}, {Bonato}, {Calistro Rivera}, {Cassano}, {Cochrane}, {Croston}, {Cuciti}, {Dallacasa}, {Danezi}, {Dettmar}, {Di Gennaro}, {Edler}, {En{\ss}lin}, {Emig}, {Franzen}, {Garc{\'\i}a-Vergara}, {Grange}, {G{\"u}rkan}, {Hajduk}, {Heald}, {Heesen}, {Hoang}, {Hoeft}, {Horellou}, {Iacobelli}, {Jamrozy}, {Jeli{\'c}}, {Kondapally}, {Kukreti}, {Kunert-Bajraszewska}, {Magliocchetti}, {Mahatma}, {Ma{\l}ek}, {Mandal}, {Massaro}, {Meyer-Zhao}, {Mingo}, {Mostert}, {Nair},
  {Nakoneczny}, {Nikiel-Wroczy{\'n}ski}, {Orr{\'u}}, {Pajdosz-{\'S}mierciak}, {Pasini}, {Prandoni}, {van Piggelen}, {Rajpurohit}, {Retana-Montenegro}, {Riseley}, {Rowlinson}, {Saxena}, {Schrijvers}, {Sweijen}, {Siewert}, {Timmerman}, {Vaccari}, {Vink}, {West}, {Wo{\l}owska}, {Zhang}, \& {Zheng}}]{Shimwell2022}
{Shimwell}, T.~W., {Hardcastle}, M.~J., {Tasse}, C., {et~al.} 2022, \aap, 659, A1

\bibitem[{{Shimwell} {et~al.}(2017){Shimwell}, {R{\"o}ttgering}, {Best}, {Williams}, {Dijkema}, {de Gasperin}, {Hardcastle}, {Heald}, {Hoang}, {Horneffer}, {Intema}, {Mahony}, {Mandal}, {Mechev}, {Morabito}, {Oonk}, {Rafferty}, {Retana-Montenegro}, {Sabater}, {Tasse}, {van Weeren}, {Br{\"u}ggen}, {Brunetti}, {Chy{\.z}y}, {Conway}, {Haverkorn}, {Jackson}, {Jarvis}, {McKean}, {Miley}, {Morganti}, {White}, {Wise}, {van Bemmel}, {Beck}, {Brienza}, {Bonafede}, {Calistro Rivera}, {Cassano}, {Clarke}, {Cseh}, {Deller}, {Drabent}, {van Driel}, {Engels}, {Falcke}, {Ferrari}, {Fr{\"o}hlich}, {Garrett}, {Harwood}, {Heesen}, {Hoeft}, {Horellou}, {Israel}, {Kapi{\'n}ska}, {Kunert-Bajraszewska}, {McKay}, {Mohan}, {Orr{\'u}}, {Pizzo}, {Prandoni}, {Schwarz}, {Shulevski}, {Sipior}, {Smith}, {Sridhar}, {Steinmetz}, {Stroe}, {Varenius}, {van der Werf}, {Zensus}, \& {Zwart}}]{Shimwell2017}
{Shimwell}, T.~W., {R{\"o}ttgering}, H.~J.~A., {Best}, P.~N., {et~al.} 2017, \aap, 598, A104

\bibitem[{{Shimwell} {et~al.}(2019){Shimwell}, {Tasse}, {Hardcastle}, {Mechev}, {Williams}, {Best}, {R{\"o}ttgering}, {Callingham}, {Dijkema}, {de Gasperin}, {Hoang}, {Hugo}, {Mirmont}, {Oonk}, {Prandoni}, {Rafferty}, {Sabater}, {Smirnov}, {van Weeren}, {White}, {Atemkeng}, {Bester}, {Bonnassieux}, {Br{\"u}ggen}, {Brunetti}, {Chy{\.z}y}, {Cochrane}, {Conway}, {Croston}, {Danezi}, {Duncan}, {Haverkorn}, {Heald}, {Iacobelli}, {Intema}, {Jackson}, {Jamrozy}, {Jarvis}, {Lakhoo}, {Mevius}, {Miley}, {Morabito}, {Morganti}, {Nisbet}, {Orr{\'u}}, {Perkins}, {Pizzo}, {Schrijvers}, {Smith}, {Vermeulen}, {Wise}, {Alegre}, {Bacon}, {van Bemmel}, {Beswick}, {Bonafede}, {Botteon}, {Bourke}, {Brienza}, {Calistro Rivera}, {Cassano}, {Clarke}, {Conselice}, {Dettmar}, {Drabent}, {Dumba}, {Emig}, {En{\ss}lin}, {Ferrari}, {Garrett}, {G{\'e}nova-Santos}, {Goyal}, {G{\"u}rkan}, {Hale}, {Harwood}, {Heesen}, {Hoeft}, {Horellou}, {Jackson}, {Kokotanekov}, {Kondapally}, {Kunert-Bajraszewska}, {Mahatma}, {Mahony}, {Mandal}, {McKean},
  {Merloni}, {Mingo}, {Miskolczi}, {Mooney}, {Nikiel-Wroczy{\'n}ski}, {O'Sullivan}, {Quinn}, {Reich}, {Roskowi{\'n}ski}, {Rowlinson}, {Savini}, {Saxena}, {Schwarz}, {Shulevski}, {Sridhar}, {Stacey}, {Urquhart}, {van der Wiel}, {Varenius}, {Webster}, \& {Wilber}}]{Shimwell2019}
{Shimwell}, T.~W., {Tasse}, C., {Hardcastle}, M.~J., {et~al.} 2019, \aap, 622, A1

\bibitem[{{Skalidis} {et~al.}(2021){Skalidis}, {Sternberg}, {Beattie}, {Pavlidou}, \& {Tassis}}]{Skalidis2021}
{Skalidis}, R., {Sternberg}, J., {Beattie}, J.~R., {Pavlidou}, V., \& {Tassis}, K. 2021, \aap, 656, A118

\bibitem[{{Skrutskie} {et~al.}(2006){Skrutskie}, {Cutri}, {Stiening}, {Weinberg}, {Schneider}, {Carpenter}, {Beichman}, {Capps}, {Chester}, {Elias}, {Huchra}, {Liebert}, {Lonsdale}, {Monet}, {Price}, {Seitzer}, {Jarrett}, {Kirkpatrick}, {Gizis}, {Howard}, {Evans}, {Fowler}, {Fullmer}, {Hurt}, {Light}, {Kopan}, {Marsh}, {McCallon}, {Tam}, {Van Dyk}, \& {Wheelock}}]{Skrutskie2006}
{Skrutskie}, M.~F., {Cutri}, R.~M., {Stiening}, R., {et~al.} 2006, \aj, 131, 1163

\bibitem[{{Smirnov} \& {Tasse}(2015)}]{Smirnov2015}
{Smirnov}, O.~M. \& {Tasse}, C. 2015, \mnras, 449, 2668

\bibitem[{{Sokolowski} {et~al.}(2022){Sokolowski}, {Tingay}, {Davidson}, {Wayth}, {Ung}, {Broderick}, {Juswardy}, {Kovaleva}, {Macario}, {Pupillo}, \& {Sutinjo}}]{Sokolowski2022}
{Sokolowski}, M., {Tingay}, S.~J., {Davidson}, D.~B., {et~al.} 2022, \pasa, 39, e015

\bibitem[{{Soler}(2019)}]{Soler2019b}
{Soler}, J.~D. 2019, \aap, 629, A96

\bibitem[{{Stanislavsky} {et~al.}(2023){Stanislavsky}, {Bubnov}, {Konovalenko}, {Stanislavsky}, \& {Yerin}}]{Stanislavsky2023}
{Stanislavsky}, L.~A., {Bubnov}, I.~N., {Konovalenko}, A.~A., {Stanislavsky}, A.~A., \& {Yerin}, S.~N. 2023, \aap, 670, A157

\bibitem[{{Tasse}(2014)}]{Tasse2014}
{Tasse}, C. 2014, \aap, 566, A127

\bibitem[{{Tasse} {et~al.}(2018){Tasse}, {Hugo}, {Mirmont}, {Smirnov}, {Atemkeng}, {Bester}, {Hardcastle}, {Lakhoo}, {Perkins}, \& {Shimwell}}]{Tasse2018}
{Tasse}, C., {Hugo}, B., {Mirmont}, M., {et~al.} 2018, \aap, 611, A87

\bibitem[{{Thompson} {et~al.}(2019){Thompson}, {Troland}, \& {Heiles}}]{Thompson2019}
{Thompson}, K.~L., {Troland}, T.~H., \& {Heiles}, C. 2019, \apj, 884, 49

\bibitem[{{Tobin} {et~al.}(2016){Tobin}, {Looney}, {Li}, {Chandler}, {Dunham}, {Segura-Cox}, {Sadavoy}, {Melis}, {Harris}, {Kratter}, \& {Perez}}]{Tobin2016}
{Tobin}, J.~J., {Looney}, L.~W., {Li}, Z.-Y., {et~al.} 2016, \apj, 818, 73

\bibitem[{{Tychoniec} {et~al.}(2018){Tychoniec}, {Tobin}, {Karska}, {Chandler}, {Dunham}, {Harris}, {Kratter}, {Li}, {Looney}, {Melis}, {P{\'e}rez}, {Sadavoy}, {Segura-Cox}, \& {van Dishoeck}}]{Tychoniec2018}
{Tychoniec}, {\L}., {Tobin}, J.~J., {Karska}, A., {et~al.} 2018, \apjs, 238, 19

\bibitem[{{van Diepen} {et~al.}(2018){van Diepen}, {Dijkema}, \& {Offringa}}]{vanDiepen2018}
{van Diepen}, G., {Dijkema}, T.~J., \& {Offringa}, A. 2018, {DPPP: Default Pre-Processing Pipeline}, Astrophysics Source Code Library, record ascl:1804.003

\bibitem[{{van Haarlem} {et~al.}(2013){van Haarlem}, {Wise}, {Gunst}, {Heald}, {McKean}, {Hessels}, {de Bruyn}, {Nijboer}, {Swinbank}, {Fallows}, {Brentjens}, {Nelles}, {Beck}, {Falcke}, {Fender}, {H{\"o}randel}, {Koopmans}, {Mann}, {Miley}, {R{\"o}ttgering}, {Stappers}, {Wijers}, {Zaroubi}, {van den Akker}, {Alexov}, {Anderson}, {Anderson}, {van Ardenne}, {Arts}, {Asgekar}, {Avruch}, {Batejat}, {B{\"a}hren}, {Bell}, {Bell}, {van Bemmel}, {Bennema}, {Bentum}, {Bernardi}, {Best}, {B{\^\i}rzan}, {Bonafede}, {Boonstra}, {Braun}, {Bregman}, {Breitling}, {van de Brink}, {Broderick}, {Broekema}, {Brouw}, {Br{\"u}ggen}, {Butcher}, {van Cappellen}, {Ciardi}, {Coenen}, {Conway}, {Coolen}, {Corstanje}, {Damstra}, {Davies}, {Deller}, {Dettmar}, {van Diepen}, {Dijkstra}, {Donker}, {Doorduin}, {Dromer}, {Drost}, {van Duin}, {Eisl{\"o}ffel}, {van Enst}, {Ferrari}, {Frieswijk}, {Gankema}, {Garrett}, {de Gasperin}, {Gerbers}, {de Geus}, {Grie{\ss}meier}, {Grit}, {Gruppen}, {Hamaker}, {Hassall}, {Hoeft}, {Holties},
  {Horneffer}, {van der Horst}, {van Houwelingen}, {Huijgen}, {Iacobelli}, {Intema}, {Jackson}, {Jelic}, {de Jong}, {Juette}, {Kant}, {Karastergiou}, {Koers}, {Kollen}, {Kondratiev}, {Kooistra}, {Koopman}, {Koster}, {Kuniyoshi}, {Kramer}, {Kuper}, {Lambropoulos}, {Law}, {van Leeuwen}, {Lemaitre}, {Loose}, {Maat}, {Macario}, {Markoff}, {Masters}, {McFadden}, {McKay-Bukowski}, {Meijering}, {Meulman}, {Mevius}, {Middelberg}, {Millenaar}, {Miller-Jones}, {Mohan}, {Mol}, {Morawietz}, {Morganti}, {Mulcahy}, {Mulder}, {Munk}, {Nieuwenhuis}, {van Nieuwpoort}, {Noordam}, {Norden}, {Noutsos}, {Offringa}, {Olofsson}, {Omar}, {Orr{\'u}}, {Overeem}, {Paas}, {Pand ey-Pommier}, {Pandey}, {Pizzo}, {Polatidis}, {Rafferty}, {Rawlings}, {Reich}, {de Reijer}, {Reitsma}, {Renting}, {Riemers}, {Rol}, {Romein}, {Roosjen}, {Ruiter}, {Scaife}, {van der Schaaf}, {Scheers}, {Schellart}, {Schoenmakers}, {Schoonderbeek}, {Serylak}, {Shulevski}, {Sluman}, {Smirnov}, {Sobey}, {Spreeuw}, {Steinmetz}, {Sterks}, {Stiepel}, {Stuurwold},
  {Tagger}, {Tang}, {Tasse}, {Thomas}, {Thoudam}, {Toribio}, {van der Tol}, {Usov}, {van Veelen}, {van der Veen}, {ter Veen}, {Verbiest}, {Vermeulen}, {Vermaas}, {Vocks}, {Vogt}, {de Vos}, {van der Wal}, {van Weeren}, {Weggemans}, {Weltevrede}, {White}, {Wijnholds}, {Wilhelmsson}, {Wucknitz}, {Yatawatta}, {Zarka}, {Zensus}, \& {van Zwieten}}]{vanHaarlem2013}
{van Haarlem}, M.~P., {Wise}, M.~W., {Gunst}, A.~W., {et~al.} 2013, \aap, 556, A2

\bibitem[{{van Weeren} {et~al.}(2016){van Weeren}, {Williams}, {Hardcastle}, {Shimwell}, {Rafferty}, {Sabater}, {Heald}, {Sridhar}, {Dijkema}, {Brunetti}, {Br{\"u}ggen}, {Andrade-Santos}, {Ogrean}, {R{\"o}ttgering}, {Dawson}, {Forman}, {de Gasperin}, {Jones}, {Miley}, {Rudnick}, {Sarazin}, {Bonafede}, {Best}, {B{\^\i}rzan}, {Cassano}, {Chy{\.z}y}, {Croston}, {Ensslin}, {Ferrari}, {Hoeft}, {Horellou}, {Jarvis}, {Kraft}, {Mevius}, {Intema}, {Murray}, {Orr{\'u}}, {Pizzo}, {Simionescu}, {Stroe}, {van der Tol}, \& {White}}]{vanWeeren2016}
{van Weeren}, R.~J., {Williams}, W.~L., {Hardcastle}, M.~J., {et~al.} 2016, \apjs, 223, 2

\bibitem[{{Virtanen} {et~al.}(2020){Virtanen}, {Gommers}, {Oliphant}, {Haberland}, {Reddy}, {Cournapeau}, {Burovski}, {Peterson}, {Weckesser}, {Bright}, {van der Walt}, {Brett}, {Wilson}, {Millman}, {Mayorov}, {Nelson}, {Jones}, {Kern}, {Larson}, {Carey}, {Polat}, {Feng}, {Moore}, {VanderPlas}, {Laxalde}, {Perktold}, {Cimrman}, {Henriksen}, {Quintero}, {Harris}, {Archibald}, {Ribeiro}, {Pedregosa}, {van Mulbregt}, \& {SciPy 1. 0 Contributors}}]{Virtanen2020}
{Virtanen}, P., {Gommers}, R., {Oliphant}, T.~E., {et~al.} 2020, Nature Methods, 17, 261

\bibitem[{{Williams} {et~al.}(2016){Williams}, {van Weeren}, {R{\"o}ttgering}, {Best}, {Dijkema}, {de Gasperin}, {Hardcastle}, {Heald}, {Prandoni}, {Sabater}, {Shimwell}, {Tasse}, {van Bemmel}, {Br{\"u}ggen}, {Brunetti}, {Conway}, {En{\ss}lin}, {Engels}, {Falcke}, {Ferrari}, {Haverkorn}, {Jackson}, {Jarvis}, {Kapi{\'n}ska}, {Mahony}, {Miley}, {Morabito}, {Morganti}, {Orr{\'u}}, {Retana-Montenegro}, {Sridhar}, {Toribio}, {White}, {Wise}, \& {Zwart}}]{Williams2016}
{Williams}, W.~L., {van Weeren}, R.~J., {R{\"o}ttgering}, H.~J.~A., {et~al.} 2016, \mnras, 460, 2385

\bibitem[{{Zucker} {et~al.}(2019){Zucker}, {Speagle}, {Schlafly}, {Green}, {Finkbeiner}, {Goodman}, \& {Alves}}]{Zucker2019}
{Zucker}, C., {Speagle}, J.~S., {Schlafly}, E.~F., {et~al.} 2019, \apj, 879, 125

\end{thebibliography}

\appendix

\section{Mosaic of five fields of view from LoTSS}\label{app:mosaic}

\begin{figure}[!h]
\begin{center}
\resizebox{0.9\hsize}{!}{\includegraphics{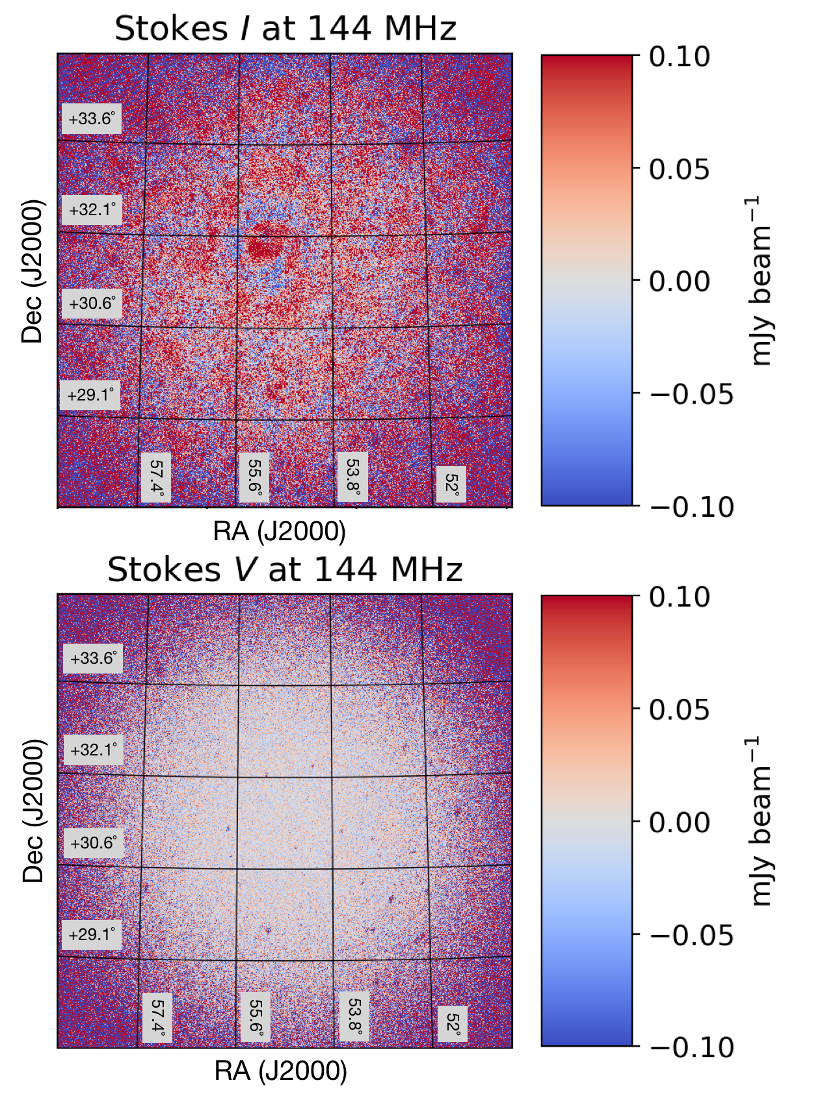}}
\caption{Stokes $I$ and $V$ maps of Perseus corresponding to the single 8h pointing P054+31.}
\label{fig:lofar1n}
\end{center}
\end{figure}

In this study, we combined five single-night 8-hour pointings from the LOFAR internal data release, referred to as P051+31, P052+29, P054+31, P057+31, and P057+34. Once prepared for publication (Shimwell et al., in prep.), the data will be accessible through the LoTSS repository\footnote{ \url{https://repository.surfsara.nl/collection/lotss-dr2}}. The mosaic was constructed in both Stokes $I$ and Stokes $V$ (see Fig.~\ref{fig:lofardata}). As an example, in Fig.~\ref{fig:lofar1n} we show the two Stokes parameter maps for one single pointing, specifically P054+31. Both maps share the same coordinate grid and projection as the $N_{\rm H_2}$ map shown in Fig.~\ref{fig:coldens0}. 

\begin{figure}[!h]
\begin{center}
\resizebox{0.9\hsize}{!}{\includegraphics{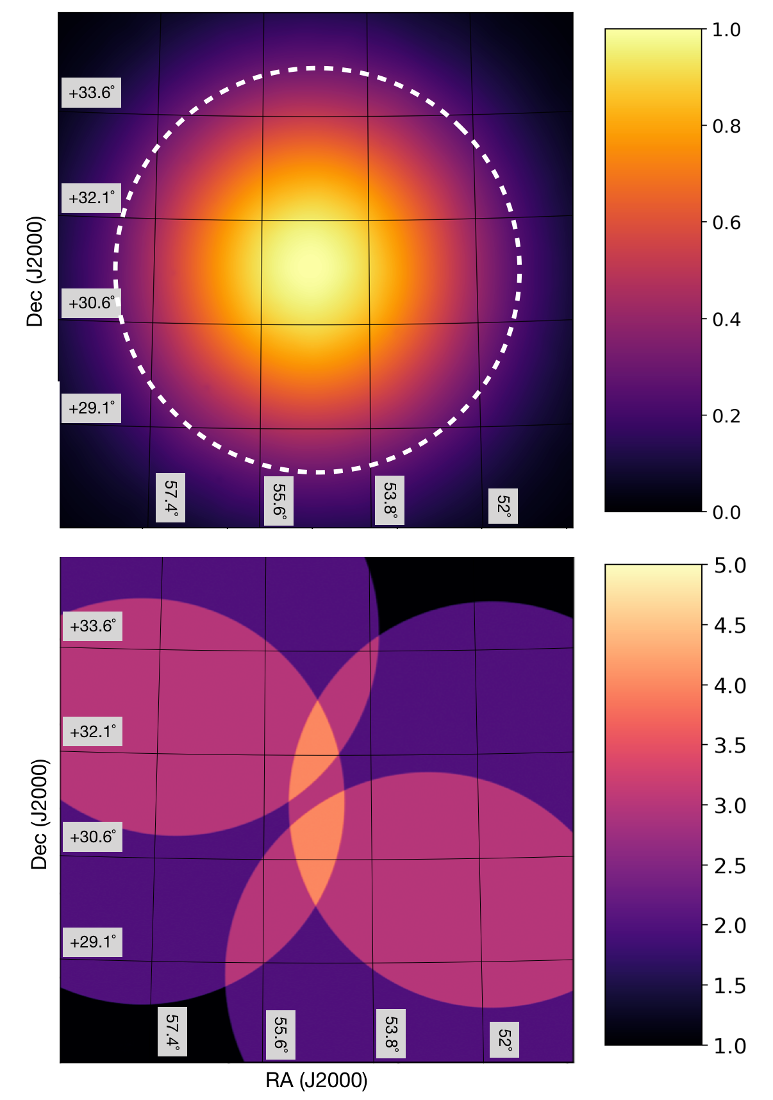}}
\caption{LOFAR primary-beam weighting (top panel) and overlap of the five pointings used for the mosaic (bottom panel). A white-dashed circle of 3-deg radius in the top panel defines the region in each pointing utilised to produce the mosaic.}
\label{fig:mosaic}
\end{center}
\end{figure}

The single-night pointing maps reveal that noise -- as traced by Stokes $V$ -- is highly non uniform across the field of view, presenting challenges for our stacking analysis. The increased noise at the edges of the image is caused by LOFAR’s primary-beam sensitivity. In the top panel of Fig.~\ref{fig:mosaic}, we show the primary-beam weighting of LOFAR with a 3-deg radius, $r_{\rm pb}$, in white-dashed contour. Beyond $r_{\rm pb}$, sensitivity drops rapidly. We combined all five pointings, including only the corresponding circular area, $\pi r_{\rm pb}^2$. The overlap of these pointings is illustrated in the bottom panel of Fig.~\ref{fig:mosaic}.     

\begin{figure}[!h]
\begin{center}
\resizebox{0.9\hsize}{!}{\includegraphics{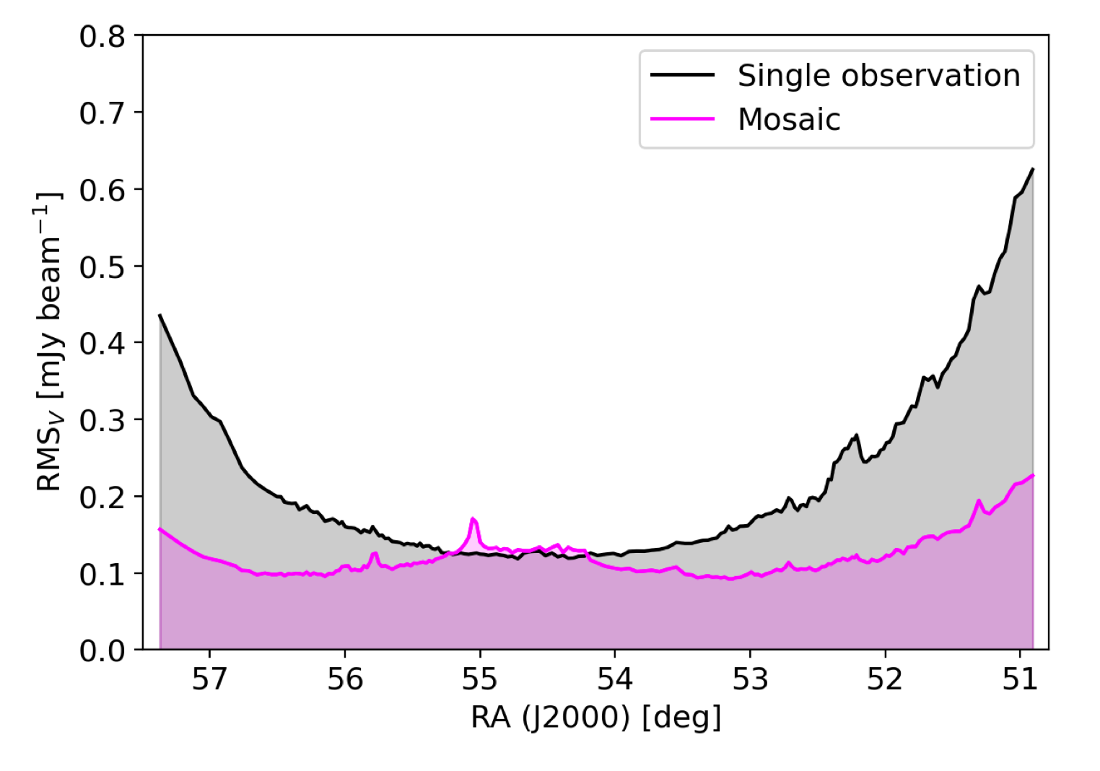}}
\caption{Root mean square (RMS) of Stokes $V$ as a function of RA for the P054+31 pointing (in black) for $N_{\rm H_2} > 10^{21}$ cm$^{-2}$ (see Fig.~\ref{fig:coldens0}). The single-night RMS is compared to the RMS of the mosaic shown in magenta.}
\label{fig:rmsV}
\end{center}
\end{figure}

The quantitative advantage of the mosaic with respect to single pointing observations is seen in the noise levels in Fig.~\ref{fig:rmsV}. Since the Perseus molecular cloud has an elongated RA extent (see Fig.~\ref{fig:coldens0}), we present the RMS value of Stokes $V$ as a function of RA for $N_{\rm H_2} > 10^{21}$ cm$^{-2}$, where most of our core samples are located (see Fig.~\ref{fig:coldens0}). For a single pointing (P054+31), the RMS varies by more than a factor of five (black line), while in the mosaic, it remains stable between 0.1 and 0.2 mJy beam$^{-1}$ (magenta line).          
\onecolumn
\section{Cut-out maps of the bright cores in LOFAR}
\label{app:cutouts}

\begin{figure*}[!ht]
\begin{center}
\resizebox{0.85\hsize}{!}{\includegraphics{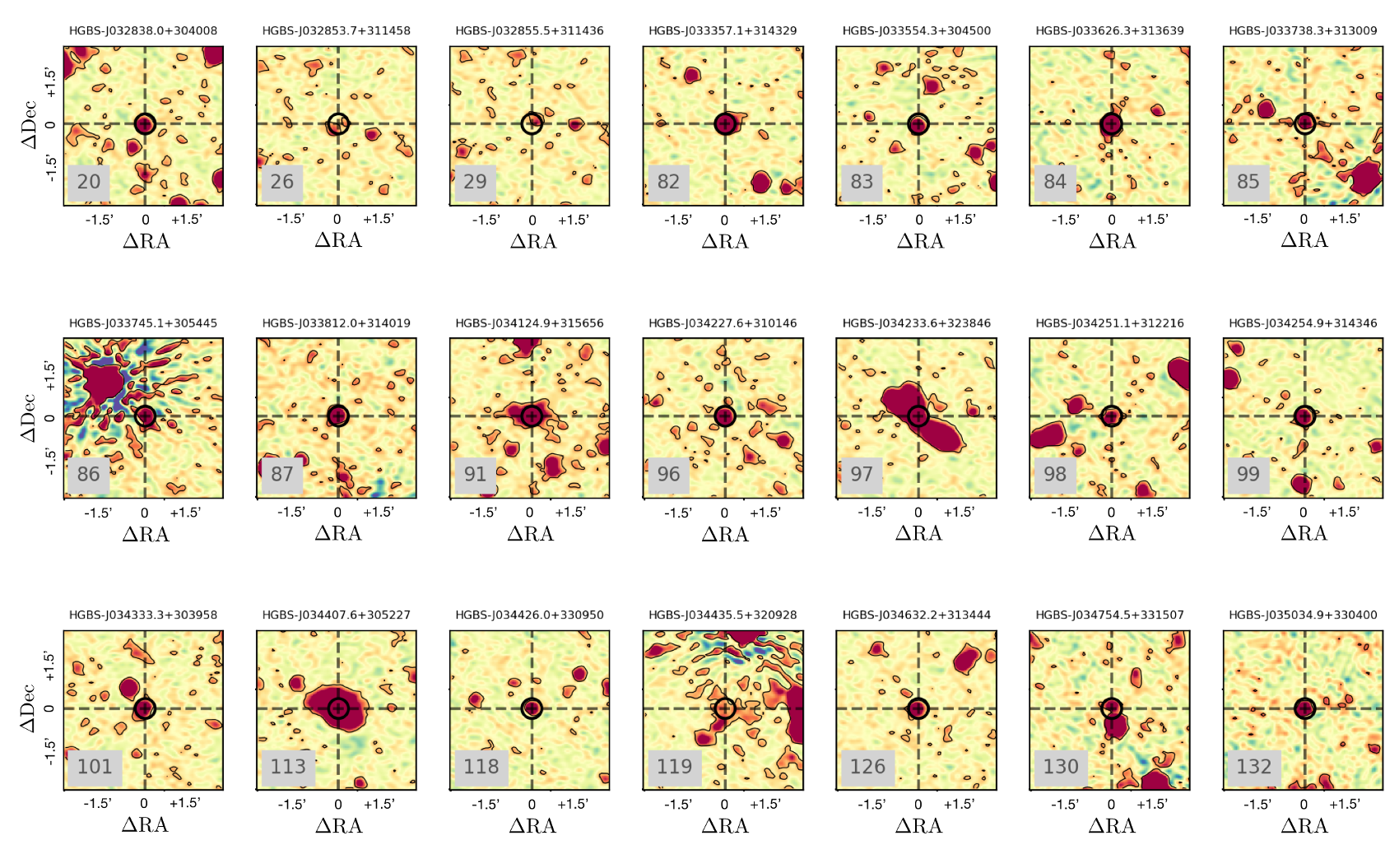}}
\caption{Stokes-$I$ cut-outs around each protostar selected in Fig.~\ref{fig:sel}. The corresponding source name as in \citetalias{Pezzuto2021} and index are displayed in gray. The color-range for all sources is between $-1$ and 1 mJy beam$^{-1}$.}
\label{fig:proto}
\end{center}
\end{figure*}

\begin{figure*}[!ht]
\begin{center}
\resizebox{0.85\hsize}{!}{\includegraphics{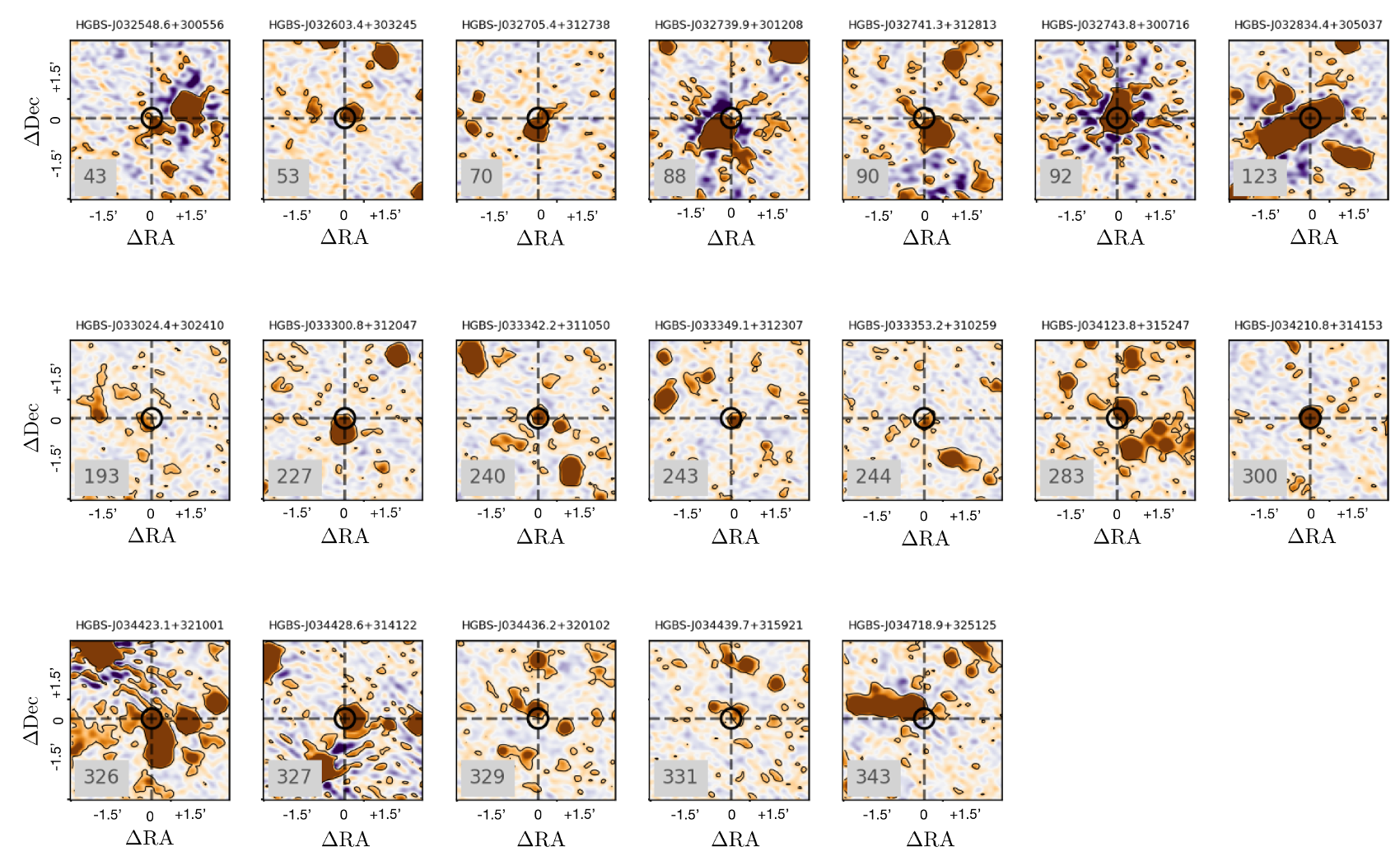}}
\caption{Same as in Fig.~\ref{fig:proto} but for prestellar cores using the same dynamic range between $-1$ and 1 mJy beam$^{-1}$.}
\label{fig:pres}
\end{center}
\end{figure*}

In this Appendix we show the cut-outs of each selected bright core in the LOFAR data, as described in Sect.~\ref{ssec:stack1}. In Figs.~\ref{fig:proto} and ~\ref{fig:pres}, the corresponding images of protostellar and prestellar cores are displayed. 

\end{document}